\documentclass[12pt]{article}

\usepackage{amsmath,amsfonts}
\usepackage{graphicx,psfrag,epsf,algorithm}
\usepackage{enumerate}
\usepackage{natbib}
\usepackage[usenames,dvipsnames,svgnames,table]{xcolor}

\newcommand{\blind}{0}

\newcommand\bigfrac[2]{\frac{\displaystyle{#1}}{\displaystyle{#2}}}

\newcommand{\Phat}{\widehat{p}}
\newcommand{\Ptil}{\widehat{p}_{a}}

\begin{document}

\def\spacingset#1{\renewcommand{\baselinestretch}%
{#1}\small\normalsize} \spacingset{1}


\if0\blind
{
  \title{Delayed acceptance particle MCMC for exact inference in
    stochastic kinetic models}
  \author{Andrew Golightly\thanks{andrew.golightly@ncl.ac.uk}\hspace{.2cm}\\
    {\small School of Mathematics \& Statistics, Newcastle University, NE1 7RU, UK}\\
    and \\
    Daniel A. Henderson \\
    {\small School of Mathematics \& Statistics, Newcastle University, NE1 7RU, UK}\\
    and \\
    Chris Sherlock \\
    {\small Department of Mathematics and Statistics, Lancaster University, LA1 4YF, UK}}

  \maketitle
} \fi

\if1\blind
{
  \bigskip
  \bigskip
  \bigskip
  \begin{center}
    {\LARGE\bf Title}
\end{center}
  \medskip
} \fi

\bigskip

\maketitle

\begin{abstract}
Recently-proposed particle MCMC methods provide a flexible way of performing 
Bayesian inference for parameters governing stochastic kinetic models defined as Markov 
(jump) processes (MJPs). Each iteration of the scheme requires an
estimate of the marginal likelihood calculated 
from the output of a sequential Monte Carlo scheme (also known as a particle filter). 
Consequently, the method can be extremely computationally intensive. 
We therefore aim to avoid most instances of the expensive likelihood calculation through 
use of a fast approximation. We consider two approximations: the
chemical Langevin equation diffusion approximation (CLE) and the
linear noise approximation (LNA). Either an estimate of the marginal likelihood 
under the CLE, or the tractable marginal likelihood under the LNA can be used 
to calculate a first step acceptance probability. Only if a proposal is accepted 
under the approximation do we then run a sequential Monte Carlo scheme to compute an estimate of the 
marginal likelihood under the true MJP and construct a second stage acceptance probability that 
permits exact (simulation based) inference for the MJP. We therefore avoid expensive 
calculations for proposals that are likely to be rejected. We illustrate the method 
by considering inference for parameters governing a Lotka-Volterra system, a model of gene expression 
and a simple epidemic process.
\end{abstract}

\noindent%

\spacingset{1.1}

\section{Introduction}
Stochastic kinetic models describe a probabilistic mechanism for the joint 
evolution of species in a dynamical system. They can be used to model a wide 
variety of real-world phenomena and are increasingly applied in computational 
systems biology \citep{kitano2002}, motivated by a need for models 
that incorporate intrinsic stochasticity \citep{elowitz2002,swain2002,wilkinson2009}. 
Other areas of application include (but are not limited to) predator-prey population 
models \citep{boys08,ferm2008,golightly11} and epidemic models \citep{oneill1999,boys2007,ball2008,jewell2009}. 
Underpinned by a reaction network in which reaction events change species numbers by an integer 
amount, a stochastic kinetic model is most naturally represented by a continuous time Markov jump process (MJP). 
Our goal is to perform inference for the rate constants that govern 
the MJP using time course data that may be incomplete and/or subject to measurement error. 

Exact (simulation based) Bayesian inference for the MJP 
was the subject of \cite{boys08}. The authors proposed two MCMC schemes that targeted 
the joint posterior of the rate constants and latent reaction events but found the statistical 
efficiency of their method to be relatively poor. It was shown in \cite{golightly11} how a recently proposed particle MCMC 
algorithm \citep{andrieu10} can be applied to this class of models. 
In particular, the particle marginal Metropolis-Hastings (PMMH) scheme allows 
a joint update of the rate constants and (latent) process which can alleviate 
common mixing problems when sampling high dimensional target densities 
that may exhibit strong correlations. The proposal mechanism involves drawing a new parameter 
value from an arbitrary proposal kernel and drawing new values of each latent state from a 
sequential Monte Carlo (SMC) approximation to the distribution of latent states conditional 
on the proposed new parameter value. The acceptance probability requires 
computation of a realisation of an unbiased estimator of marginal 
likelihood which can be readily obtained from the output of the SMC scheme. Consequently, at 
each iteration of the MH scheme, an SMC algorithm must be implemented. The method can 
be extremely computationally intensive, as the SMC algorithm typically must generate many realisations of the MJP, with each
  realisation obtained from an algorithm such as the stochastic simulation algorithm (SSA) of \cite{Gillespie77}. By using 
a computationally cheaper approximation to the marginal likelihood we avoid running the computationally more expensive 
SMC algorithm at most iterations of the MH scheme, but we still maintain the posterior under the MJP as the target distribution of the 
MH scheme. 

The simplest approximation of the MJP is the macroscopic rate equation (MRE) which ignores 
the discreteness and stochasticity of the MJP by modelling specie dynamics 
with a set of coupled ordinary differential equations \citep{kampen2001}. The diffusion 
approximation or chemical Langevin equation (CLE) \citep{Gillespie2000} on the other hand, ignores discreteness but not stochasticity by 
modelling the reaction network with a set of coupled stochastic differential 
equations (SDEs). Whilst inference for the parameters governing nonlinear multivariate 
SDEs is possible \citep{Golightly08}, the marginal likelihood under 
this model is intractable. Despite this, \cite{golightly11} show that inference is possible 
under this model using a PMMH algorithm, and this approach can result in computational savings when compared 
to a similar scheme targeting the posterior under the MJP. 

Further computational savings can be made by considering a linear noise approximation (LNA) 
\citep{kampen2001,Komorowski09,fearnhead12} which is given by the MRE plus a stochastic term accounting for 
random fluctuations about the MRE. Under the LNA the latent process follows a multivariate 
Gaussian distribution and, under an assumption of Gaussian measurement error, the marginal 
likelihood is tractable.    
   
\cite{ChristenF05} describe a delayed-acceptance Metropolis-Hastings scheme in which the single 
MH accept-reject step is replaced by an initial `screening' stage which substitutes a computationally 
cheap approximate posterior for the true posterior in the MH acceptance probability formula, 
but then adds a second accept-reject stage which ensures that detailed balance is still 
satisfied with respect to the true posterior. This second, computationally expensive, 
stage is only applied to proposals which pass the first stage.

Our novel contribution is to exploit the tractability of the LNA by proposing a particle analogue 
of this scheme for performing exact, simulation based inference for the MJP parameters. Essentially, 
to avoid calculating an estimate of marginal likelihood 
under the MJP for proposals that are likely to be rejected, proposed parameter draws are 
initially screened using a computationally cheap approximation to the posterior, such as that based on the marginal likelihood 
computed under the LNA. A related approach has been proposed independently by \cite{Smith11} for performing inference 
for the parameters governing nonlinear, discrete time economic models. A simple 
stochastic volatility model and a Real Business Cycle model are considered, with approximations based on a linear Gaussian 
state space model and an unscented Kalman filter used in a preliminary screening step. Unlike \cite{Smith11}, we also consider 
a scenario in which the marginal likelihood under the approximation is intractable, but can be 
estimated cheaply (relative to the same calculation under the MJP) using a particle filter. Use of the 
CLE in the preliminary screening step falls into this category. In both cases, we show that the 
resulting MCMC scheme targets the correct marginal, that is, the marginal parameter posterior 
under the MJP. The proposed methods can in principle be applied to any Markov jump process.  

The remainder of this paper is organised as follows. In
Section~\ref{skm} we describe the Markov jump process model and associated
inference problem. The CLE and LNA are briefly reviewed. We describe
the PMMH algorithm in Section~\ref{pmmh} before considering 
a modification to allow delayed acceptance in Section~\ref{em}. We 
apply the method to a Lotka-Volterra system, a model of gene expression and a simple epidemic process in Section~\ref{app}. 
Conclusions are drawn in Section~\ref{conc}.

\section{Stochastic kinetic models}\label{skm}
Consider a reaction network involving $u$ species $\mathcal{X}_1,\mathcal{X}_2,\linebreak[1]
\ldots,\mathcal{X}_u$ and $v$ reactions $\mathcal{R}_1,\mathcal{R}_2,\ldots,\mathcal{R}_v$, 
with reaction $\mathcal{R}_i$ given by
\begin{align*}
& \mathcal{R}_i:\quad p_{i1}\mathcal{X}_1+p_{i2}\mathcal{X}_2+\cdots+p_{iu}\mathcal{X}_u \\ 
& \hspace{2cm} \longrightarrow q_{i1}\mathcal{X}_1+q_{i2}\mathcal{X}_2+\cdots+q_{iu}\mathcal{X}_u 
\end{align*}
where the stoichiometric coefficients $p_{ij}$ and $q_{ij}$ are non-negative integers. 
Let $X_{j,t}$ denote the number of specie $\mathcal{X}_j$ at time
$t$, and let $X_t$ be the $u$-vector $X_t =
(X_{1,t},X_{2,t},\linebreak[1] \ldots,\linebreak[0] X_{u,t})'$. The $v\times u$ matrix $P$ consists of
the coefficients $p_{ij}$, and $Q$ is defined similarly. The $u\times v$
stoichiometry matrix $S$ is defined by
\[
S = (Q-P)'
\]
and encodes important structural
information about the reaction network. In particular,
if $\Delta R$ is a $v$-vector containing the
number of reaction events of each type in a given time interval, then
the system state should be updated by $\Delta X$, where
\[
\Delta X = S \Delta R.
\] 
Each reaction $\mathcal{R}_i$ is assumed to have an associated rate
constant, $c_i$, and a propensity function, $h_i(X_t,c_i)$
giving the overall hazard of a type $i$ reaction
occurring. That is, we model the system as a Markov jump process, and
for an infinitesimal time increment $dt$, the probability of a type
$i$ reaction occurring in the time interval $(t,t+dt]$ is
$h_i(X_t,c_i)dt$. In many examples (such as those considered in Sections~\ref{lv} and \ref{epi}) 
the form of $h_i(X_t,c_i)$ can be thought of as arising 
naturally from the interactions between components of a well-mixed population, 
such as reactants in a well-stirred container at constant temperature. This leads to a 
mass action kinetic rate law \citep{gillespie1992}, under which 
the hazard function for a particular 
reaction of type $i$ takes the form 
\[
h_i(X_t,c_i) = c_i\prod_{j=1}^u \binom{X_{j,t}}{p_{ij}}.
\]
Let $c=(c_1,c_2,\ldots,c_v)'$ and
$h(X_t,c)=(h_1(X_t,c_1),\linebreak[1]h_2(X_t,c_2),\ldots,h_v(X_t,c_v))'$. Values for $c$ and the initial system state $X_0=x_0$
complete specification of the Markov process. Although this process is
rarely analytically tractable for interesting models, it is
straightforward to forward-simulate exact realisations of this Markov
process using a discrete event simulation method. This is due to the
fact that if the current time and state of the system are $t$ and
$X_t$ respectively, then the time to the next event will be
exponential with rate parameter
\[
h_0(X_t,c)=\sum_{i=1}^v h_i(X_t,c_i),
\]
and the event will be a reaction of type $\mathcal{R}_i$ with probability
$h_i(X_t,c_i)/h_0(X_t,c)$ independently of the waiting time. Forward
simulation of process realisations in this way is typically referred
to as the stochastic simulation algorithm \citep{Gillespie77}. 
See \cite{Wilkinson12} for further background on stochastic kinetic modelling.

\subsection{Chemical Langevin equation}

We present here an informal intuitive construction of the chemical Langevin 
equation (CLE), and 
refer the reader to \cite{Gillespie2000} for further details.

Consider an infinitesimal time interval, $(t,t+dt]$. Over this time, the
reaction hazards will remain constant almost surely. The
occurrence of reaction events can therefore be regarded as the
occurrence of events of a Poisson process with independent realisations
for each reaction type. Therefore, if we write $dR_t$ for the
$v$-vector of the number of reaction events of each type in the infinitesimal time
increment, it is clear that the elements are independent of one
another and that the $i$th element is a $Po(h_i(X_t,c_i)dt)$ random
quantity. From this we have that $\textrm{E}(dR_t)=h(X_t,c)dt$ and
$\textrm{Var}(dR_t)=\textrm{diag}\{h(X_t,c)\}dt$. It is
therefore clear that
\[
dR_t = h(X_t,c)dt + \textrm{diag}\left\{\sqrt{h(X_t,c)}\right\}dW_t
\]
is the It\^o stochastic differential equation (SDE) which has the same
infinitesimal mean and variance as the true Markov jump process (where
$dW_t$ is the increment of a $v$-dimensional Brownian motion). Now
since $dX_t=S dR_t$, we obtain
\begin{equation}\label{cle}
d X_t = S\,h(X_t,c)dt + \sqrt{S\textrm{diag}\{h(X_t,c)\}S'}dW_t,
\end{equation}
where now $X_t$ and $W_t$ are both $u$-vectors. Equation (\ref{cle})
is the SDE most commonly referred to as the chemical Langevin
equation or diffusion approximation, and represents the diffusion process which most
closely matches the dynamics of the associated Markov jump process, 
and can be shown to approximate the stochastic kinetic model increasingly well in high
concentration scenarios \citep{Gillespie2000}. Note that in the absence 
of an analytic solution to (\ref{cle}), a numerical solution can be constructed. 
For example, the Euler-Maruyama approximation is
\begin{align}\label{cle-em}
\Delta X_{t}&\equiv  X_{t+\Delta t}-X_{t}\nonumber\\
&= S\,h(X_t,c)\Delta t + \sqrt{S\textrm{diag}\{h(X_t,c)\}S'}\Delta W_t
\end{align}  
where $\Delta W_t$ is a mean zero Normal random vector with variance matrix 
$\textrm{diag}\{\Delta t\}$. 

We require a computationally efficient approximation 
to the Markov jump process for use in a delayed acceptance particle MCMC scheme 
(described in Section~\ref{em}). Performing exact (simulation based) 
inference for the diffusion approximation has been the focus of 
\cite{Golightly05}, \cite{Purutcuoglu07}, 
and \cite{golightly11} among others. Although the latter find that a particle 
MCMC scheme based on the CLE can be more computationally 
efficient than a similar scheme that works with the Markov jump process directly, 
calculation of an estimate of marginal likelihood under the CLE (as is necessary 
at every iteration of a particle MCMC scheme) can be computationally expensive. 
To facilitate greater computational savings, we therefore also consider 
a linear noise approximation (LNA) 
\citep{kampen2001,Komorowski09,fearnhead12,stathopoulos13} which generally possesses a greater degree 
of numerical and analytic tractability than the CLE \citep{Wilkinson12}. 
This is the subject of the next section.

\subsection{Linear noise approximation}
The LNA was first considered as a functional central limit law for density dependent processes 
by \cite{kurtz1970} and can be derived in a number of more or less formal ways. For example, 
\cite{Komorowski09} (and see also \cite{elf2003}) derive the LNA by approximating the 
forward Kolmogorov equation (satisfied by the transition rate of the MJP) through a Taylor 
series expansion. We eschew this approach in favour of an informal derivation 
following that of \cite{fearnhead12} and we 
refer the reader to the references therein for a more detailed discussion. In what follows 
we calculate the LNA for a general SDE before formulating it 
as an approximation to the CLE.

Consider now a general SDE satisfied by a process $\{X_{t},t\geq 0\}$ of the form
\begin{equation}\label{genSDE}
dX_{t}=\alpha(X_t)dt + \epsilon\beta(X_t)dW_t
\end{equation}
where $\epsilon <<1$. Partition $X_t$ into a deterministic path $z_t$ and a residual 
stochastic process $M_t$ and let $z_{t}$ be the solution to
\begin{equation}\label{ev1}
\frac{dz_{t}}{dt}=\alpha(z_{t}).
\end{equation}  
We assume that $||X_t-z_t||$ is $O(\epsilon)$ over a time interval of interest and substitute 
$X_{t}=z_t+\epsilon M_t$ into equation~(\ref{genSDE}) to give
\[
d(z_t+\epsilon M_t) =\alpha(z_t+\epsilon M_t)dt + \epsilon\beta(z_t+\epsilon M_t)dW_t.
\]
We then Taylor expand $\alpha(\cdot)$ and $\beta(\cdot)$ about $z_t$ and collect 
terms of $O(\epsilon)$ to give the SDE satisfied by $M_t$ as 
\begin{equation}\label{ev2}
dM_{t}= F_t M_t dt + \beta(z_t)dW_t 
\end{equation}
where $F_t$ is the Jacobian matrix with $(i,j)$th 
element $\partial \alpha_{i}(z_t) / \partial z_{j,t}$ and $\alpha_{i}(z_t)$ refers to the 
$i$th element of $\alpha(z_t)$. 

We use $\epsilon$ to explicitly indicate that the stochastic term in (\ref{genSDE}) is small. 
Its presence helps us to gather together terms that are small but not negligible 
(i.e. $O(\epsilon))$. We may instead remove the explicit $\epsilon$ 
(effectively setting $\epsilon=1$) and simply think of $\beta(X_t)$ 
as small. Since $\epsilon$ plays no role in the evolution equations, 
(\ref{ev1}) and (\ref{ev2}), these are unchanged whether we define 
$M_t$ as $(X_t-z_t)/\epsilon$ or as $X_t-z_t$; only the initial 
condition for (\ref{ev2}) and the interpretation of $M_t$ change 
since now $M_t=X_t-z_t$. Without loss of generality, 
therefore, we simplify the exposition by setting $\epsilon=1$ 
and assuming that $\beta(X_t)$ itself is small. To further simplify 
the notation we also drop the explicit dependence of the hazard function on $c$, and of the mean and 
variance of $M_t$ on both $c$ and $z_t$. 

For the CLE, we have 
\[
\alpha(X_t)= S\,h(X_t),\qquad \beta(X_t)=\sqrt{S\textrm{diag}\{h(X_t)\}S'}.
\]
The linear noise approximation of the CLE is therefore defined through
\begin{equation}\label{LNA1}
\frac{dz_{t}}{dt}= Sh(z_t)
\end{equation} 
and
\begin{equation}\label{LNA2}
dM_{t}= F_t M_t dt +\sqrt{S\textrm{diag}\{h(z_t)\}S'} dW_t 
\end{equation}
where $F_t$ has $(i,j)$th 
element given by the first partial derivative of the $i$th element of $S\,h(z_t)$ with respect to $z_{j,t}$.

For fixed or Gaussian initial conditions, that is 
$M_{t_{1}}\sim \textrm{N}(m_{t_{1}},V_{t_{1}})$,  the SDE in (\ref{LNA2}) can be solved 
explicitly to give
\begin{equation}\label{mtd}
\left(M_{t}|c\right) \sim \textrm{N}\left(m_{t}\,,\,V_{t}\right)
\end{equation} 
where $m_{t}$ is the solution to the deterministic ordinary differential equation (ODE)
\begin{equation}\label{odeG}
\frac{dm_{t}}{dt} = F_t m_{t}
\end{equation}
and similarly
\begin{equation}\label{odeV}
\frac{dV_{t}}{dt} = V_{t}F_{t}' + S\textrm{diag}\{h(z_t)\}S' + F_{t}V_{t} \,.
\end{equation}
Hence, the solution of equation~(\ref{LNA2}) requires the solution of 
a system of coupled ODEs; in the absence of an 
analytic solution to these equations, a numerical solution can be used. The approximating distribution 
of $X_{t}$ can then be found as
\begin{equation}
\left(X_{t}|c\right) \sim \textrm{N}\left(z_{t}+m_{t}\,,\, V_{t}\right).
\end{equation}  

\section{Inference}\label{inf}
We now consider the task of performing inference for the rate constants governing the 
Markov jump process. First, let us augment the rate vector $c$ to include any additional 
parameters that arise from the observation process and assign to it a prior density, $p(c)$. 
Suppose that the MJP $\mathbf{X}=\{X_{t}\,|\, 1\leq t \leq T\}$ is not observed directly, but (perhaps partial) 
observations (on a regular grid) $\mathbf{y}=\{y_{t}\,|\, t=1,2,\ldots ,T\}$ are available 
and assumed conditionally independent (given $\mathbf{X}$) with conditional probability 
distribution $p(y_{t}|x_{t},c)$. 

In this work, we consider Bayesian inference for $c$ via the marginal posterior density
\begin{equation}\label{jp}
p(c|\mathbf{y}) = \int p(c,\mathbf{x}|\mathbf{y})\,\textrm{d}\mathbf{x}
\end{equation}
where
\[
p(c,\mathbf{x}|\mathbf{y})\propto p(c)\,p(\mathbf{x}|c)\,\prod_{t=1}^{T}p(y_{t}|x_{t},c)
\]
and $p(\mathbf{x}|c)$ is the 
probability of the Markov jump process. Since the 
posterior in (\ref{jp}) will typically be unavailable in closed form, samples 
must usually be generated through a suitable MCMC scheme.

In what follows, for simplicity, we assume that the initial value of the MJP, $X_1 = x_1$, is 
a known fixed quantity, and we take $z_1 = x_1$ so that $m_1$ is the length-$u$ zero vector and 
$V_1$ is the $u\times u$ zero matrix. If $X_1$ were unknown then it could be assigned a prior and 
treated as an additional parameter in the augmented rate vector.

\subsection{Particle marginal Metropolis-Hastings}\label{pmmh}
We consider the special case of the particle marginal Metropolis-Hastings (PMMH) scheme of \cite{andrieu10} 
and \cite{andrieu09} in which only samples from the marginal parameter posterior are required. 
Noting the standard decomposition $p(c|\mathbf{y})\propto p(\mathbf{y}|c)p(c)$, 
we run a Metropolis-Hastings (MH) scheme with proposal kernel $q(c^{*}|c)$ and accept 
a move from $c$ to $c^{*}$ with probability
\begin{equation}\label{aratio2}
\textrm{min}\left\{1\,,\,\frac{\widehat{p}\big(\mathbf{y}|c^{\star}\big)p\big(c^{\star}\big)}{\widehat{p}\big(\mathbf{y}|c\big)p\big(c\big)} \times \frac{q\big(c|c^{\star}\big)}
{q\big(c^{\star}|c\big)}\right\}
\end{equation}
where $\widehat{p}(\mathbf{y}|c)$ is a sequential Monte Carlo (SMC) or `particle filter' estimate of 
the intractable marginal likelihood term $p(\mathbf{y}|c)$. The PMMH scheme as described here is 
an example of a pseudo-marginal Metropolis-Hastings scheme \citep{beaumont03,andrieu09b} 
and provided that $\widehat{p}(\mathbf{y}|c)$ is unbiased (or has a
constant multiplicative bias that does not depend on $c$), it is possible to verify that the method targets the marginal 
$p(c |\mathbf{y})$. Let $u$ denote all random variables generated by the SMC algorithm 
and write the SMC estimate of marginal likelihood as $\widehat{p}(\mathbf{y}|c)=\widehat{p}(\mathbf{y}|c,u)$. 
Augmenting the state space of the chain to include $u$, it is straightforward to rewrite 
the acceptance ratio in (\ref{aratio2}) to find that the chain targets the joint density  
\[
\widehat{p}(c,u|\mathbf{y})\propto \widehat{p}(\mathbf{y}|c,u)\tilde{q}(u|c)p(c)
\]
where $\tilde{q}(u|c)$ denotes the conditional density associated with the auxiliary variables $u$. 
Marginalising over $u$ then gives
\begin{align*}
\int \widehat{p}(c,u|\mathbf{y})du &\propto p(c)\int \widehat{p}(\mathbf{y}|c,u)\tilde{q}(u|c)du\\
&\propto p(c) p(\mathbf{y}|c)\,.
\end{align*}
The key insight here is that the SMC scheme can be constructed to give an unbiased estimate of the marginal 
likelihood $p(\mathbf{y}|c)$ under some fairly mild conditions involving the resampling scheme 
\citep{delmoral04}. The scheme therefore targets the correct marginal $p(c |\mathbf{y})$. 
Although interest here is in the marginal $p(c
|\mathbf{y})$ the PMMH scheme 
can be used to sample the joint density $p(c,\mathbf{x}|\mathbf{y})$. At each step of the algorithm, 
a new path $\mathbf{x}^{*}$ is proposed from an SMC approximation of $p(\mathbf{x}^{*}|\mathbf{y},c^{*})$. 
The acceptance probability is as in (\ref{aratio2}). For further details, we refer the reader 
to \cite{andrieu10}. The (special case of the) PMMH algorithm and details of the SMC scheme that we use 
are given in Appendices~\ref{A} and \ref{A1}.

\subsection{Inference using the CLE and LNA}\label{LNAinf}
Although the marginal likelihood under the CLE is intractable, a PMMH scheme 
can be implemented to perform inference for this model. In the simplest version 
of the scheme, we replace 
draws of the MJP in step 2(a) of the SMC scheme with draws of a numerical 
solution of the CLE, for example, using the Euler-Maruyama approximation. This is 
the focus of \cite{golightly11} and further details can be found therein.

Under the LNA, the marginal likelihood is tractable for additive Gaussian observation regimes. 
This tractability has been exploited 
for the purposes of parameter inference by \cite{Komorowski09}, \cite{fearnhead12} and 
\cite{stathopoulos13}. In \cite{Komorowski09} and \cite{stathopoulos13}, the LNA is applied over the entire time interval for which 
observations are available. In particular, the ODE component of the LNA is solved once 
over the whole time-course for a given initial condition. As discussed in \cite{fearnhead12}, 
this can lead to a poor approximation to the distribution of $X_t$ as $t$ gets large, 
due to the mismatch between the stochastic and ODE solution. We therefore 
adopt the approach proposed in \cite{fearnhead12} and restart the LNA at each observation 
time $t$, initialising $z_t$ to the posterior mean of $X_{t}$ given all observations 
up to time $t$. The 
algorithm for constructing the marginal likelihood under an additive Gaussian observation regime 
using this approach is given in Appendix~\ref{A2}. Use of a Gaussian observation model is likely to be 
unsatisfactory in some scenarios. For example, in Section~\ref{lv} we consider observations with a Poisson 
distribution, the mean of which is the value of the true process. Nonetheless, we may still use the LNA 
to obtain a tractable approximation to the marginal likelihood under the true MJP. We approximate the observation 
density $p(y_t|x_{t})$ by a Gaussian density with mean and variance given by the ODE solution (\ref{LNA1}). That is, 
we apply the algorithm in Appendix~\ref{A2} with $\Sigma$ replaced by a diagonal matrix containing the components 
of $z_t$ for which observations are made. This tractable approximation can then be used in the delayed acceptance 
scheme. 

\subsection{Delayed acceptance  particle marginal Metropolis-Hastings}\label{em}
In order to improve the efficiency of the PMMH algorithm for the MJP we aim to
limit the number of runs of the computationally expensive SMC scheme
for the MJP. Ideally we want to run the SMC scheme only for parameter
values which are likely to lead to acceptance in the PMMH
algorithm. We do this by choosing a particular proposal kernel in the
PMMH scheme of Appendix A.1.  This proposal kernel is based on a
preliminary screening step involving an approximate model which is
less computationally intensive than the MJP, such as the
LNA or the CLE. In what follows, the CLE approximation refers to the Euler-Maruyama 
approximation in (\ref{cle-em}). Likewise, the LNA refers to the numerical solution 
of the ODEs in (\ref{LNA1}), (\ref{odeG}) and (\ref{odeV}). We note that the CLE or LNA are used only 
in the preliminary screening step and further approximation through use of a numerical solution will not 
change the target distribution of the Metropolis-Hastings scheme. 

Our proposed algorithm for taking advantage of the CLE approximation, which we call delayed acceptance PMMH
(daPMMH), is outlined in Algorithm~\ref{alg:dapmmh}; the
  algorithm which takes advantage of the LNA is a slight
  simplification of this. Both algorithms have the
following basic structure.  

\begin{algorithm}[]
\caption{Delayed acceptance PMMH (daPMMH)}
\label{alg:dapmmh}
\begin{enumerate}[1.]
\item Initialisation, $i=0$,
\begin{enumerate}[(a)]
\item set $c^{(0)}$ arbitrarily,
\item run a particle filter targeting $p(\mathbf{x}|\mathbf{y},c^{(0)})$,
  and let $\widehat{p}(\mathbf{y}|c^{(0)})$ denote the marginal
  likelihood estimate,
\item run a particle filter targeting $p_a(\mathbf{x}|\mathbf{y},c^{(0)})$,
  and let $\widehat{p}_a(\mathbf{y}|c^{(0)})$ denote the marginal
  likelihood estimate under the approximate model.
\end{enumerate}
\item For iteration $i\geq 1$, 
\begin{enumerate}[(a)]
\item sample $c^*\sim q(\cdot |c^{(i-1)})$,
\item \textbf{Stage 1}
\begin{enumerate}[(i)]
\item run a particle filter targeting $p_a(\mathbf{x}|\mathbf{y},c^*)$,
  and let $\widehat{p}_a(\mathbf{y}|c^*)$ denote the marginal
  likelihood estimate under the approximate model,
\item with probability 
\begin{eqnarray}
\alpha_1(c^{(i-1)},c^*)&=&
\min\left\{1,\bigfrac{\widehat{p}_a(\mathbf{y}|c^*)p(c^*)}{\widehat{p}_a(\mathbf{y}|c^{(i-1)})p(c^{(i-1)})}\bigfrac{q(c^{(i-1)}|c^*)}{q(c^*|c^{(i-1)})}\right\},
\label{eq:acc-cheap-smc}
\end{eqnarray}
run a particle filter targeting
$p(\mathbf{x}|\mathbf{y},c^*)$, let $\widehat{p}(\mathbf{y}|c^*)$ denote
the marginal likelihood estimate and go to 2(c); 
otherwise,  set $c^{(i)}=c^{(i-1)}$,
$\widehat{p}(\mathbf{y}|c^{(i)})=\widehat{p}(\mathbf{y}|c^{(i-1)})$, 
$\widehat{p}_a(\mathbf{y}|c^{(i)})=\widehat{p}_a(\mathbf{y}|c^{(i-1)})$, increment $i$ and return to 2(a).
\end{enumerate}

\item \textbf{Stage 2}

With probability 
\begin{eqnarray}
\alpha_2(c^{(i-1)},c^*)&=& \min\left\{1,\bigfrac{\widehat{p}(\mathbf{y}|c^*)p(c^*)}{\widehat{p}(\mathbf{y}|c^{(i-1)})p(c^{(i-1)})}\bigfrac{\widehat{p}_a(\mathbf{y}|c^{(i-1)})p(c^{(i-1)})}{\widehat{p}_a(\mathbf{y}|c^*)p(c^*)}\right\}
\label{eq:acc-exact-smc}
\end{eqnarray}
 set $c^{(i)}=c^*$, $\widehat{p}(\mathbf{y}|c^{(i)})=\widehat{p}(\mathbf{y}|c^*)$ and
$\widehat{p}_a(\mathbf{y}|c^{(i)})=\widehat{p}_a(\mathbf{y}|c^*)$ otherwise set
$c^{(i)}=c^{(i-1)}$, $\widehat{p}(\mathbf{y}|c^{(i)})=\widehat{p}(\mathbf{y}|c^{(i-1)})$ and
$\widehat{p}_a(\mathbf{y}|c^{(i)})=\widehat{p}_a(\mathbf{y}|c^{(i-1)})$. Increment $i$ and return to 2(a).
\end{enumerate}
\end{enumerate}
\end{algorithm}

First a candidate set of parameter values is proposed, then a decision
is made whether to accept or reject these values based on a MH step
with target density $p_a(c|\mathbf{y})\propto p_a(\mathbf{y}|c)p(c)$, which
is the posterior density of parameters under the approximate model
(for example, the LNA or the CLE); here $p_a(\mathbf{y}|c)$ represents
the marginal likelihood under the approximate model. If the proposed
parameter values are accepted at this first stage then they undergo
another MH step with target density $p(c|\mathbf{y})\propto
p(\mathbf{y}|c)p(c)$, which is the marginal posterior density under the
MJP. The idea here is that the first stage weeds out `poor' parameter
values. Consequently, the computationally expensive SMC algorithm for
the MJP is only implemented for `good' parameter values which are
likely to be accepted at the second stage.  

When the CLE is used as the approximate model the marginal likelihood
$p_a(\mathbf{y}|c)$ is not available analytically, so we
replace it with an unbiased estimate $\widehat{p}_a(\mathbf{y}|c)$
obtained from an SMC scheme which targets $p_a(\mathbf{x}|\mathbf{y},c)$,
the conditional density of the
latent states under the approximate model, given the observed data and
the parameter values. We therefore have to run a particle filter at
both stages of the daPMMH algorithm, as one is always needed at stage 2 to
give an unbiased estimate $\widehat{p}(\mathbf{y}|c)$ of the MJP marginal
likelihood $p(\mathbf{y}|c)$. We note, however, that despite
the CLE requiring a run of an SMC scheme to obtain
$\widehat{p}_a(\mathbf{y}|c)$ this may still be much faster to run than
the SMC scheme for the MJP (with the same number of particles).

Our daPMMH algorithm is an extension of the delayed acceptance MH
(daMH) algorithm of \cite{ChristenF05}, which is a version of the
`surrogate transition method' of \cite{Liu01}. Specifically, we have
extended the daMH algorithm by replacing all intractable marginal
likelihoods by unbiased estimates obtained from appropriate SMC
schemes. Our extension of the daMH algorithm to an intractable
likelihood at Stage 1 is essential when the approximate model is the
CLE since the marginal likelihood under the CLE is intractable.
However, when the LNA is chosen as the approximate model this extra
level of complexity is not necessary; we simply replace the marginal
likelihood estimates $\widehat{p}_a(\mathbf{y}|c)$ in
Algorithm~\ref{alg:dapmmh} with the exact values $p_a(\mathbf{y}|c)$
since these are available numerically (see Appendix A.3 for details).
Despite replacing the intractable marginal likelihoods by unbiased
estimates, our daPMMH algorithm still targets the (exact) posterior
density of the parameters under the MJP, $p(c|\mathbf{y})$, as we
outline in Section~\ref{sec:dapmmh}. Note that in an independent
technical report, \cite{Smith11} proved that the daPMMH algorithm has
$p(c|\mathbf{y})$ as its target density when the marginal likelihood
under the approximate model is tractable.  In Section~\ref{sec:dapmmh}
we generalise the argument of \cite{Smith11} to the case of an
SMC-based marginal likelihood estimate for the approximate model.

\subsubsection{Validity of delayed acceptance PMMH}
\label{sec:dapmmh}

In this section we show that the daPMMH algorithm
(Algorithm~\ref{alg:dapmmh}) is a valid MCMC scheme which targets a
distribution that admits $p(c|\mathbf{y})$ as a marginal distribution.

We first define some notation and an extended state-space. 
Let $F:\mathbb{R}^2\rightarrow [0,1]$ be any function satisfying the following.
\begin{eqnarray}
\label{eqn.F.prop.A}
aF[a,a^*]&=&a^*F[a^*,a]\\
\label{eqn.F.prop.B}
F[ba,ba^*]&=&F[a,a^*].
\end{eqnarray}
An example of $F$ is the Metropolis-Hastings acceptance probability
$F[a,a^*]=\min(1,a^*/a)$, with $a=p(c|\mathbf{y})q(c^*|c)$ and
$a^*=p(c^*|\mathbf{y})q(c|c^*)$. More generally, $F$ defines an acceptance
probability that admits a chain with invariant density $a$,
 a \textit{joint density} (known up to an arbitrary constant) \textit{on the current value in the chain and the next
 proposal}. Condition (\ref{eqn.F.prop.A}) ensures that
detailed balance is satisfied with respect to $a$, and Condition (\ref{eqn.F.prop.B})
ensures that the target density need only be known up to a fixed constant.

Let $U$ be a vector of auxiliary random
variables, sampled conditional on the parameters according to
$\tilde{q}(u|c)$, and let 
$\Phat(c,u)$ and $\Ptil(c,u)$
be two approximations to the posterior which depend on $U$, with $\Phat$ unbiased up
to a fixed constant, $k>0$:
\begin{equation}
\label{eqn.psm.unb}
\int \Phat(c,u)~\tilde{q}(u|c)~du=k~p(c|\mathbf{y}).
\end{equation}
Note that for notational simplicity, we have dropped dependence of $\Phat(c,u)$ and 
$\Ptil(c,u)$ on the data $\mathbf{y}$. For further clarity of exposition we adopt the shorthand
\begin{align*}
\Phat:=\Phat(c,u),&~
\Phat^*:=\Phat(c^*,u^*),~
\Ptil:=\Ptil(c,u),\\
\Ptil^*:=\Ptil(c^*,u^*),&~
\tilde{q}:=\tilde{q}(u|c),~
\tilde{q}^{*}=\tilde{q}(u^*|c^*).
\end{align*}

Our delayed-acceptance Markov chain proposes according to
$q(c^*|c)\tilde{q}^*$ and accepts with a probability of
\[
\alpha\left(c,u;c^*,u^*\right)
=
F\left[
  \Ptil~ q(c^*|c),\Ptil^* q(c|c^*)\right]
\times
F\left[
  \frac{\Phat}{\Ptil},
  \frac{\Phat^*}{\Ptil^*}
\right].
\]
Our chain targets the joint posterior $\Phat(c,u)\tilde{q}(u|c)$ so that, by
\eqref{eqn.psm.unb}, the marginal distribution for $c$ is the
posterior $p(c|\mathbf{y})$. To show that $\Phat(c,u)\tilde{q}(u|c)$ is indeed the
invariant distribution of the chain it is sufficient to show that our
chain satisfies detailed balance with respect to this posterior. Since
rejection moves ($c^*\leftarrow c,~u^*\leftarrow u$) automatically
satisfy detailed balance we need only consider moves where the
proposal is accepted.
Now 
\[
\Phat~\tilde{q}~q(c^*|c)~\tilde{q}^*
=
\Ptil~q(c^*|c)
\times \frac{\Phat~\tilde{q}~\tilde{q}^*}{\Ptil}.
\]
By \eqref{eqn.F.prop.A},
\begin{align*}
\Ptil~q(c^*|c)~F\left[
  \Ptil~ q(c^*|c),\Ptil^*~ q(c|c^*)\right] &= \Ptil^*~q(c|c^*)~F\left[\Ptil^*~ q(c|c^*),\Ptil~ q(c^*|c)\right].
\end{align*}
Also, by \eqref{eqn.F.prop.B} then \eqref{eqn.F.prop.A} then
\eqref{eqn.F.prop.B} again,
\begin{align*}
\frac{\Phat~\tilde{q}~\tilde{q}^*}{\Ptil}\times F\left[
  \frac{\Phat}{\Ptil},
  \frac{\Phat^*}{\Ptil^*}
\right]&=
\frac{\Phat~\tilde{q}~\tilde{q}^*}{\Ptil}\times 
F\left[
  \frac{\Phat~\tilde{q}~\tilde{q}^*}{\Ptil},
  \frac{\Phat^*~\tilde{q}~\tilde{q}^*}{\Ptil^*}
\right]\\
&=
\frac{\Phat^*~\tilde{q}~\tilde{q}^*}{\Ptil^*}\times 
F\left[
  \frac{\Phat^*~\tilde{q}~\tilde{q}^*}{\Ptil^*},
  \frac{\Phat~\tilde{q}~\tilde{q}^*}{\Ptil}
\right]\\
&=
\frac{\Phat^*~\tilde{q}~\tilde{q}^*}{\Ptil^*}\times 
F\left[
  \frac{\Phat^*}{\Ptil^*},
  \frac{\Phat}{\Ptil}
\right].
\end{align*}
Thus
\begin{align*}
\Phat~ \tilde{q}~q(c^*|c)~\tilde{q}^*~\alpha(c,u;c^*,u^*) &=
\Phat^*~ \tilde{q}^*~ q(c|c^*)~\tilde{q}~\alpha(c^*,u^*;c,u),
\end{align*} 
as required. 
When our Stage 1 approximation is deterministic (using the LNA ) then it is
independent of $U$. Otherwise, when we use the CLE at Stage 1, our two estimates of
the posterior are
independent, i.e. $U$ is split into two
independent vectors, $U_1$ and $U_2$, with $\Ptil$ a function of $U_1$
only and $\Phat$ a function of $U_2$ only. However, for the
algorithm to work we only need to be able to simulate $U_1$ (for Stage 1)
and then, if required, $U_2|U_1=u_1$ (for Stage 2); the independence is
not necessary. Indeed a higher Stage 2 acceptance 
rate might be obtainable if it were possible to make $\Ptil(c,U)$ and
$\Phat(c,U)$ positively correlated. Unfortunately we cannot see any 
obvious method for constructing correlated estimators
based upon the
CLE and the MJP.

\subsubsection{Comments on efficiency}
\cite{ChristenF05} note that with a fast approximate model daMH
algorithms are less computationally expensive --- that is, they
exhibit lower CPU times for the same number of iterations --- than
standard MH algorithms that do not employ delayed acceptance.  They
also note that daMH algorithms are less statistically efficient than
standard MH algorithms that do not employ delayed acceptance. Here
statistical efficiency relates to the mixing of the Markov chain, and
can be measured by the effective sample size (ESS), the number of
independent samples that are equivalent in information content to the
actual number of dependent samples from the Markov chain.  Clearly,
computational time is dictated by the speed with which $p_a(\mathbf{y}|c)$ (or its
estimate $\widehat{p}_a(\mathbf{y}|c)$) is computed, and statistical
efficiency is dictated by the accuracy of the approximation $p_a(\mathbf{y}|c)$
or $\widehat{p}_a(\mathbf{y}|c)$ to $p(\mathbf{y}|c)$. For example, $p_a(\mathbf{y}|c)$ under the LNA will
be faster to compute than $\widehat{p}_a(\mathbf{y}|c)$ under the CLE
since the latter requires a run of an SMC algorithm. However, we might
expect the CLE (at least with a small Euler time-step) to provide a better approximation to the MJP than the
LNA, since the LNA is, in some sense, a simplified version of the CLE. 
Increasing the time-step $\Delta t$ in the CLE will decrease the computation time but 
should also decrease the accuracy of the approximation; the trade-off 
in terms of computational efficiency between these two factors merits 
further investigation.

Another factor which will affect statistical efficiency is the
variability associated with the SMC-based estimate of marginal
likelihood $\widehat{p}_a(\mathbf{y}|c)$. An algorithm using
$\widehat{p}_a(\mathbf{y}|c)$ will be less statistically efficient than
an idealised algorithm which uses $p_a(\mathbf{y}|c)$ (for the same
approximate model). We might expect, therefore, that using the LNA as
the approximate model, with its tractable marginal likelihood, may
lead to increased statistical efficiency over the CLE-based
approximation, although this depends on the accuracy of the LNA.

The daPMMH scheme (using either the LNA or CLE) requires specification 
of a number of particles $N$ to be used in the SMC scheme at Stage 2. 
As noted by \cite{andrieu09b}, the mixing efficiency of the PMMH 
scheme decreases as the variance of the estimated marginal likelihood 
increases. This problem can be alleviated at the expense of greater 
computational cost by increasing $N$. This therefore suggests an optimal 
value of $N$ and finding this choice is the subject of \cite{pitt12} 
and \cite{doucet13}. The latter suggest that $N$ should be chosen so that the 
variance in the noise in the estimated log-posterior is around 1. \cite{pitt12} note that 
the penalty is small for a value between 0.25 and 2.25. We therefore recommend 
performing an initial pilot run of daPMMH to obtain an estimate of the 
posterior mean for the parameters $c$, denoted $\hat{c}$. The value 
of $N$ should then be chosen so that $\textrm{Var}(\log p(\mathbf{y}|\hat{c}))$ 
is around 1--1.5. When the CLE is used as a surrogate model, we must also specify 
a number of particles (say $N_{1}$) to be used in Stage 1. For simplicity, we 
take $N_{1}=N$. Provided the CLE is a reasonable approximation to 
the MJP, we may expect that $N_1$ provides a suitable trade-off between 
computational cost and accuracy (in terms of the variance of the estimated 
marginal likelihood under the CLE).

In the next section we show empirically that our daPMMH algorithm
(with either the CLE or the LNA as the approximate model) can lead to
improvements in overall computational efficiency (in terms of ESS
normalised by CPU time) over a vanilla PMMH scheme for the MJP.

\section{Applications}\label{app}
\subsection{Lotka-Volterra}\label{lv}
Following \cite{boys08}, we consider first a Lotka-Volterra model of predator and prey interaction comprising 
three reactions:
\begin{align*}
\mathcal{R}_1:\quad \mathcal{X}_{1} &\xrightarrow{\phantom{a}c_{1}\phantom{a}} 2\mathcal{X}_{1}\\
\mathcal{R}_2:\quad \mathcal{X}_{1}+\mathcal{X}_{2} &\xrightarrow{\phantom{a}c_{2}\phantom{a}} 2\mathcal{X}_{2}\\
\mathcal{R}_3:\quad \mathcal{X}_{2} &\xrightarrow{\phantom{a}c_{3}\phantom{a}} \emptyset.
\end{align*}
For simplicity of notation we drop the explicit dependence of the current state $X=(X_{1},X_{2})'$ 
and the deterministic approximation $z=(z_1,z_2)'$ on time, $t$. The 
stoichiometry matrix is given by
\[
S = \left(\begin{array}{rrr} 
1 & -1 & 0\\
0 & 1 & -1
\end{array}\right)
\]
and the associated hazard function is 
\[
h(X,c) = (c_{1}X_{1}, c_{2}X_{1}X_{2}, c_{3}X_{2})'.
\]
The diffusion approximation can be calculated by substituting 
$S$ and $h(X,c)$ into the CLE \eqref{cle} 
to give respective drift and diffusion coefficients of
\[
\alpha(X,c)= \begin{pmatrix}
				c_{1}X_{1}-c_{2}X_{1}X_{2} \\
				c_{2}X_{1}X_{2}-c_{3}X_{2}
			\end{pmatrix}\,,
\]
\[
\beta(X,c) = \begin{pmatrix}
			c_{1}X_{1}+c_{2}X_{1}X_{2} & -c_{2}X_{1}X_{2} \\
			 -c_{2}X_{1}X_{2}	 & c_{2}X_{1}X_{2}+c_{3}X_{2}
			\end{pmatrix}. 
\]
For the linear noise approximation, the Jacobian 
matrix $F_t$ is given by
\[
F_{t} = \left(\begin{array}{cc} 
c_{1}-c_{2}z_{2} & -c_{2}z_{1} \\
c_{2}z_{2} & c_{2}z_{1}-c_{3} 
\end{array}\right).
\]

We simulated a synthetic dataset by generating 50 observations at integer 
times using the Gillespie algorithm with initial conditions $x_{1}=(70,80)'$ 
and parameter values $c=(1.0,0.005,0.6)'$ taken from \cite{Wilkinson12}. Predator 
values were discarded leaving 50 observations on prey only. These were then corrupted 
via an error distribution for which the marginal likelihood
 under the LNA is intractable:
\[
Y_{t}\sim \textrm{Poisson}(x_{1,t}),\qquad t=1,2,\ldots ,50.
\] 
A tractable approximation to the true marginal likelihood under the MJP, for use in Stage 1 
of the delayed acceptance scheme was obtained using the LNA as described in Section~\ref{LNAinf}. 
In what follows, for simplicity, we assume that the 
latent initial state $x_{1}$ is known.

For brevity, we refer to the MCMC algorithm targeting the posterior under the MJP 
that uses the LNA inside the delayed acceptance PMMH scheme as \emph{daPMMH-LNA}. 
Similarly, when using the CLE inside the delayed acceptance scheme we refer to 
this as \emph{daPMMH-CLE}. Finally, we designate the vanilla PMMH scheme without 
delayed rejection as \emph{PMMH}. Using independent Uniform $U(-8,8)$ priors 
for each $\log(c_{i})$ we performed a pilot run of the 
PMMH scheme with 50 particles to give an approximate covariance matrix $\widehat{\textrm{Var}}(c)$ 
and approximate posterior mean $\hat{c}$. Further pilot runs were then implemented with $c$ 
fixed at $\hat{c}$ and numbers of particles ranging from 50 to 250. We found that using 
200 particles gave the variance in the noise in the estimated log-posterior as 1.16. We 
therefore took $N=200$ particles for the main monitoring runs, which consisted of $2\times 10^5$ iterations 
of each scheme, with the $\log(c_{i})$ updated in a single block using a Gaussian random walk 
proposal kernel. For PMMH, we followed the practical advice of \cite{sherlock2013} and used an innovation 
variance matrix given by $\lambda\frac{2.38^{2}}{3}\widehat{\textrm{Var}}(c)$ with $\lambda$ tuned to give an acceptance 
rate of around $10\%$. We tried a range of $\lambda$ values and report results for $\lambda=0.7$ which gave an acceptance rate of $9.4\%$.
For daPMMH-CLE and daPMMH-LNA, we found that using $\lambda=1$ and $\lambda=3$ (respectively) gave an improved overall efficiency 
(compared with simply using $\lambda=0.7$). Intuitively, as computation of an estimate of marginal likelihood under the CLE and an 
approximation to the marginal likelihood under the LNA is extremely cheap relative to the MJP, larger moves should be tried at Stage 1. 
For daPMMH-CLE, we considered three 
levels of discretisation, namely, $\Delta t=0.2, 0.125, 0.0625$. The cost of computing either an estimate of marginal likelihood 
(under the CLE) or an approximation to the marginal likelihood (under the LNA) scales roughly as $1:20:30:58:362$ for 
LNA : CLE($\Delta t=0.2$) : CLE($\Delta t=0.125$) : CLE($\Delta t\linebreak[1]=0.0625$) : MJP. All algorithms are coded in C and were 
run on a desktop computer with a 2.83 GHz clock speed.

\begin{figure*}
\centering
\psfrag{log(c1)}[][][1.2][0]{$\log(c_1)$}
\psfrag{log(c2)}[][][1.2][0]{$\log(c_2)$}
\psfrag{log(c3)}[][][1.2][0]{$\log(c_3)$}
\includegraphics[angle=270,width=\textwidth]{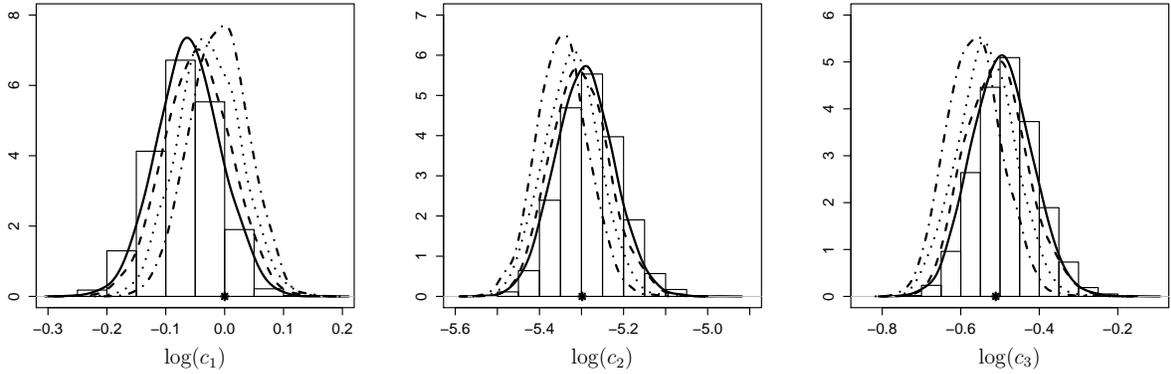}
\caption{Lotka-Volterra model. Marginal posterior distributions under the MJP (histogram), LNA (solid line) and CLE 
with $\Delta t=0.0625$ (dashed line), $\Delta t=0.125$ (dotted line),  $\Delta t=0.2$ (dot-dashed line). 
True values of each $\log(c_i)$ are indicated ($\ast$).}\label{fig:figLV}
\end{figure*}

\begin{figure*}
\centering
\psfrag{(a)}[][][1.2][0]{(a)}
\psfrag{(b)}[][][1.2][0]{(b)}
\psfrag{(c)}[][][1.2][0]{(c)}
\psfrag{(d)}[][][1.2][0]{(d)}
\psfrag{Lohpa(D1|c)}[][][1.0][0]{$\log(\widehat{p}_{a}(\mathbf{y}|c))$}
\psfrag{Logpa(D1|c)}[][][1.0][0]{$\log(p_{a}(\mathbf{y}|c))$}
\psfrag{Logp(D1|c)}[][][1.0][0]{$\log(\widehat{p}(\mathbf{y}|c))$}
\includegraphics[width=7.5cm,height=15cm,angle=270]{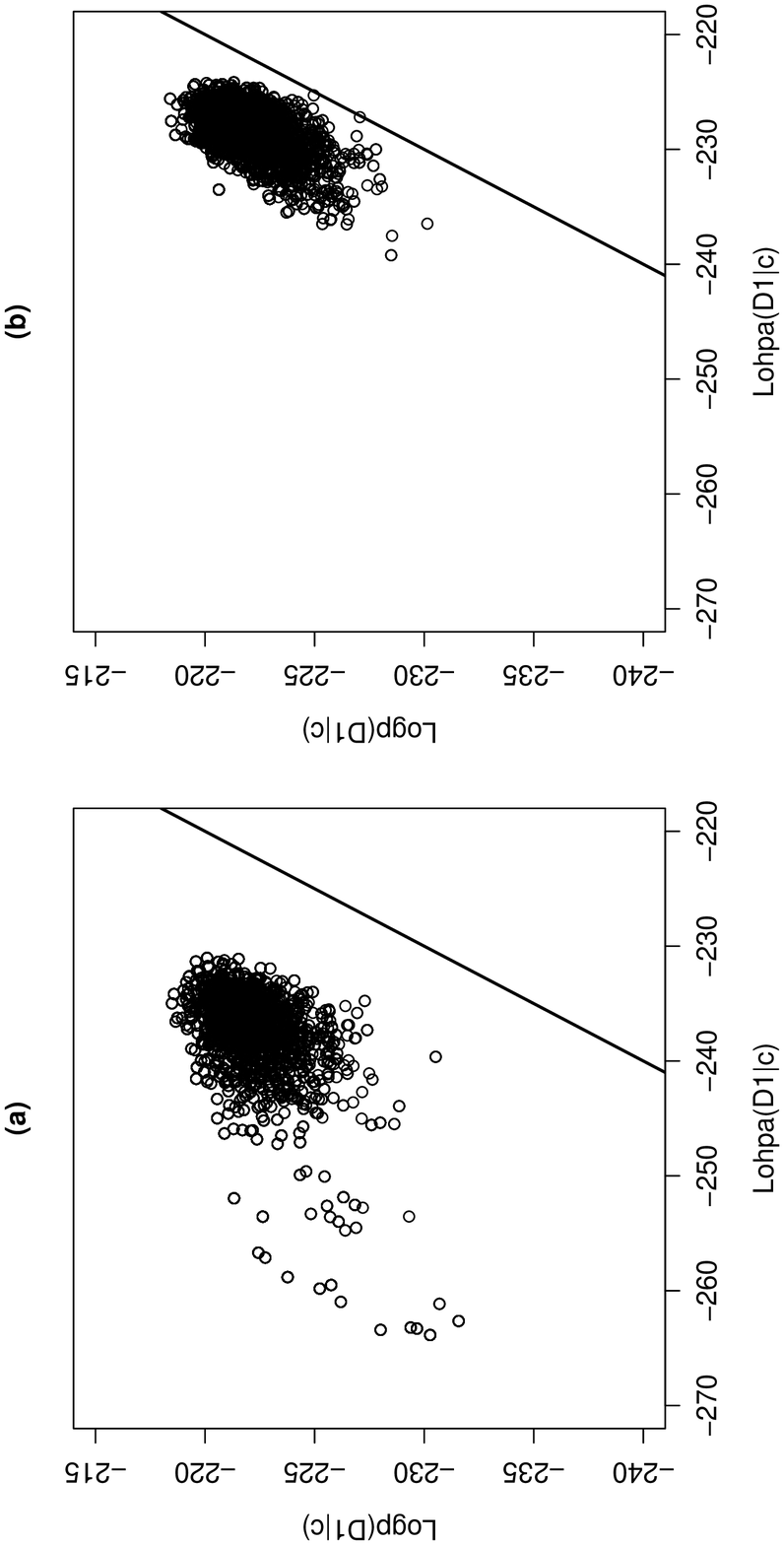}
\includegraphics[width=7.5cm,height=15cm,angle=270]{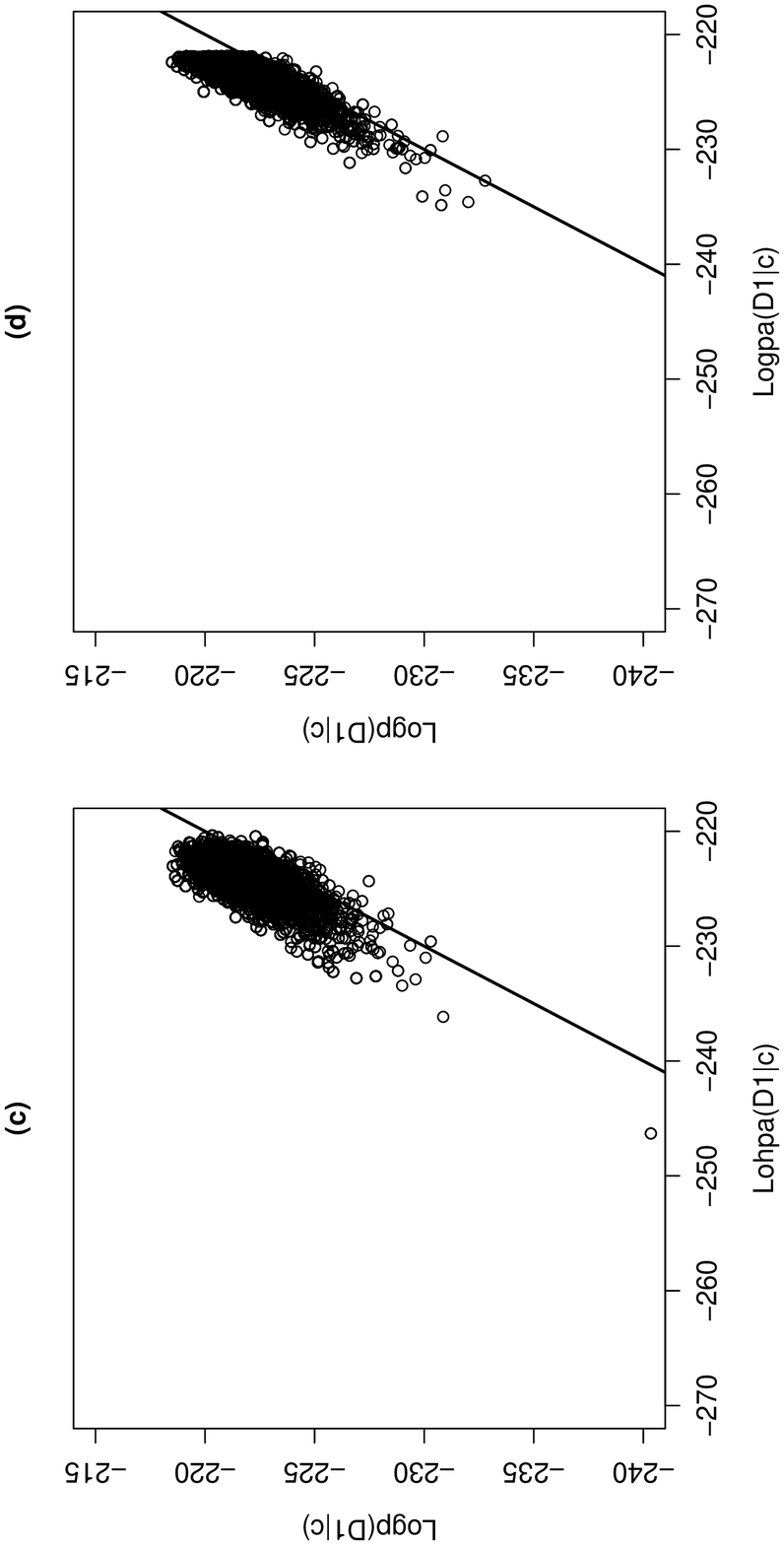}
\caption{Log-marginal likelihood estimates under the MJP ($\log(\widehat{p}(\mathbf{y}|c))$) against 
the corresponding log-marginal likelihood estimate under (a) the CLE ($\Delta t=0.2$), (b) the CLE ($\Delta t=0.125$), (c) the CLE ($\Delta t=0.0625$) 
and (d) the LNA. All plots are obtained using $10,000$ values of $c$ 
sampled from the posterior $p(c|\mathbf{y})$ for the Lotka-Volterra model.}\label{fig:fig1}
\end{figure*}

\begin{table*}[t]
	\begin{center}
	\begin{tabular}{|l|ccccc|}
	\hline	
Algorithm & $\alpha_{1}$ & $\alpha_{2|1}$ & CPU time (s) & ESS$_{min}$ &  Rel. ESS$_{min}$/s \\
\hline
PMMH                          &0.094 &1.000 &74850&2186&1.00    \\
daPMMH-CLE ($\Delta t=0.2$)   &0.123 &0.142 &15167&485 &1.10    \\
daPMMH-CLE ($\Delta t=0.125$) &0.105 &0.278 &14814&867 &2.00   \\
daPMMH-CLE ($\Delta t=0.0625$)&0.109 &0.327 &21230&948 &1.53    \\
daPMMH-LNA                    &0.031 &0.464 &2581 &835 &11.08  \\           
	\hline
	\end{tabular}
	\caption{Lotka-Volterra model. Stage 1 acceptance rate $\alpha_{1}$, Stage 2 acceptance rate $\alpha_{2|1}$, 
CPU time (to the nearest second), minimum effective sample size (ESS$_{min}$, to the nearest whole number) and 
minimum effective sample size per second, relative to the corresponding value obtained from the vanilla PMMH scheme. All values are 
based on $10^5$ iterations.}\label{tab:tab1}
	\end{center}
\end{table*}

Figure~\ref{fig:figLV} summarises the output of the PMMH scheme (consistent with the output of the delayed acceptance schemes, not reported). 
We also give kernel density estimates of the marginal parameter posteriors under the LNA and CLE (for each discretisation choice). That is, 
we ran daPMMH-LNA and daPMMH-CLE without performing the Stage 2 correction. When working with the CLE, smaller Euler time steps appear to 
give a better approximation. The effect of this choice on overall efficiency can be seen in Table~\ref{tab:tab1}. Here, we report 
Stage 1 acceptance rate $\alpha_{1}$, Stage 2 acceptance rate~$\alpha_{2|1}$, 
the CPU time, the minimum (over the 3 parameters) effective sample size (ESS$_{min}$) and minimum effective sample size per second, 
relative to the corresponding value obtained from the vanilla PMMH scheme. Whilst the daPMMH-CLE scheme 
gives an improvement in overall efficiency (as measured by relative ESS$_{min}$ per second) for 
all values of $\Delta t$ employed, the effect of the discretisation is clear. The marginal likelihood 
under the CLE approaches that under the MJP as $\Delta t$ decreases, resulting in greater statistical 
efficiency of the daPMMH-CLE scheme. This can also be seen by inspecting the Stage 2 acceptance 
probability reported in Table~\ref{tab:tab1}. Naturally, this improvement comes at a greater computational 
cost suggesting an optimal value of $\Delta t$ between 0.2 and 0.0625 for this example. Perhaps counter-intuitively, the 
CPU time for $\Delta t=0.2$ is actually greater than that for $\Delta t=0.125$. Whilst all three approximate posteriors 
that are derived from the CLE are wider than that derived from the MJP, the approximate posterior with $\Delta t=0.2$ is 
by far the widest. Consequently the Stage 1 acceptance rate is much higher and the computationally intensive Stage 2 
calculation is performed more often. Further insight into this 
result can be gained from 
Figure~\ref{fig:fig1}, which plots estimates of the marginal likelihood (on the log-scale) under PMMH 
against the corresponding value obtained under each approximation, for $10,000$ values of $c$ 
sampled from the posterior $p(c|\mathbf{y})$. The Stage 1 and 2 acceptance rates depend only on the estimates 
of the log-likelihood at the proposed and current values through their difference. Thus the efficiency 
of the algorithm is unaffected by any fixed shift of the points from the line through the origin with a slope 
of one. However, variability about a line with this slope is important and we see greater variability in the 
estimates obtained for $\Delta t=0.2$ resulting in a reduction in statistical efficiency for the 
daPMMH-CLE ($\Delta t=0.2$) scheme, with proposed values that were accepted at Stage 1 being rejected 
at Stage 2. 

The daPMMH-LNA scheme on the other hand requires minimal tuning. The LNA gives an analytic form 
for the (approximate) marginal likelihood and therefore does not require implementation 
of a particle filter during the first Stage of the delayed acceptance scheme. Moreover, the LNA solution 
involves solving a set of ODEs, for which standard routines, such as the \texttt{lsoda} package 
\citep{petzold83}, exist. Therefore, pre-specification of a suitable time discretisation is not required. 
We find for this example that the daPMMH-LNA scheme outperforms the vanilla PMMH scheme by a factor 
of more than 10. In what follows, we focus on the daPMMH-LNA scheme.     

\subsection{Gene Expression}
Here, we consider a simple model of gene expression involving three biochemical species 
(DNA, mRNA, protein) and four reaction channels (transcription, mRNA degradation, translation, 
protein degradation):
\begin{align*}
\mathcal{R}_1:\quad DNA &\xrightarrow{\phantom{a}\kappa_{R,t}\phantom{a}} DNA + R\\
\mathcal{R}_2:\quad R &\xrightarrow{\phantom{a}\gamma_{R}\phantom{a}} \emptyset\\
\mathcal{R}_3:\quad R &\xrightarrow{\phantom{a}\kappa_{P}\phantom{a}} R+P\\
\mathcal{R}_4:\quad P &\xrightarrow{\phantom{a}\gamma_{P}\phantom{a}} \emptyset.
\end{align*}    
This system has been analysed by \cite{Komorowski09} among others, and we therefore 
adopt the same notation to aid the exposition. 

Let $X_{t}=(R_{t},P_{t})'$ denote the system state at time $t$, where $R_{t}$ and 
$P_{t}$ are the respective number of mRNA and protein molecules. As in \cite{Komorowski09}, 
we take $\kappa_{R,t}$ to be the time dependent transcription rate of the gene. Specifically,
\[
\kappa_{R,t} = b_{0}\exp\left(-b_{1}(t-b_{2})^{2}\right)+b_{3}
\]
so that transcription rate increases for $t<b_{2}$ and tends to the baseline $b_{3}$ 
for $t>b_{2}$. We denote the vector of unknown parameters by 
\[
c=(\gamma_{R},\gamma_{P},\kappa_{P},b_{0},b_{1},b_{2},b_{3})'
\]
and our goal is to perform inference for these parameters. The 
stoichiometry matrix associated with the system is given by
\[
S = \left(\begin{array}{rrrr} 
1 & -1 & 0 & 0\\
0 & 0 & 1 & -1
\end{array}\right)
\]
and the associated hazard function is 
\[
h(X_{t},c) = (\kappa_{R,t}, \gamma_{R}R_{t}, \kappa_{P}R_{t}, \gamma_{P}P_{t})'.
\] 
For the linear noise approximation, we have the Jacobian matrix as
\[
F_{t} = \left(\begin{array}{cc} 
-\gamma_{R} & 0  \\
\kappa_{P}  & -\gamma_{P} 
\end{array}\right).
\] 

\begin{figure*}
\centering
\psfrag{t}[][][1.2][0]{$t$}
\psfrag{Rt}[][][1.2][270]{$R_{t}$}
\psfrag{Pt}[][][1.2][270]{$P_{t}$}
\includegraphics[width=7.5cm,height=15cm,angle=270]{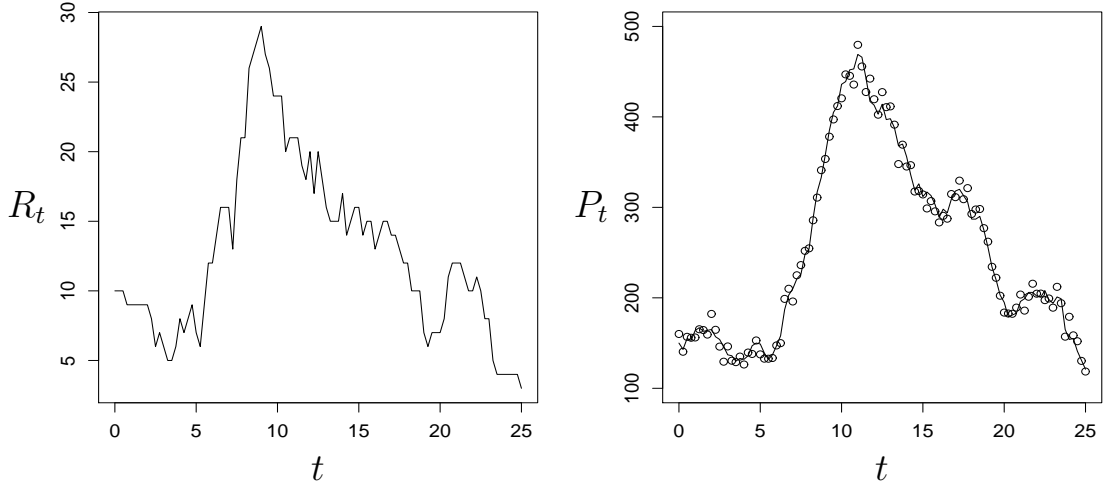}
\caption{A single realisation of the gene expression system obtained using the first reaction method. Protein 
numbers used in the artificial dataset are shown as circles.}\label{fig:fig2}
\end{figure*}

We simulated a synthetic dataset by generating observations every 15 minutes 
for 25 hours (giving 100 observations in total) noting that care must be taken 
when simulating from the MJP representation of this system, due to the time dependent 
hazard of reaction $\mathcal{R}_{1}$. We used initial conditions of $x_{1}=(10,150)'$ 
and parameter values $c=(0.44,0.52,10,15,0.4,7,3)'$ with units of time in hours. 
As in \cite{Komorowski09} we created a challenging
data-poor scenario by discarding observations on mRNA
levels and corrupting the remaining protein observations
 with additive Gaussian noise:
\[
Y_{t}\sim \textrm{N}(P_{t}\,,\,\sigma^{2}),\qquad t=1,2,\ldots ,100.
\] 
We took $\sigma=10$ and assume that this quantity is unknown. We therefore augment the 
parameter vector $c$ to include $\sigma$. The data are shown in Figure~\ref{fig:fig2}. 

For each rate constant, we assumed the same prior distributions as in \cite{Komorowski09} including 
informative priors for the degradation rates to ensure 
identifiability. Specifically, we have that
\[
\begin{array}{ll}
\gamma_{R}\sim \Gamma(19.36,44), & \gamma_{P}\sim \Gamma(27.04,52),\\
\kappa_{P}\sim \textrm{Exp}(0.01), & b_{0}\sim \textrm{Exp}(0.01),\\
b_{1}\sim \textrm{Exp}(1), & b_{2}\sim \textrm{Exp}(0.1),\\
b_{3}\sim \textrm{Exp}(0.01), & \sigma\sim \textrm{Exp}(0.01) 
\end{array}
\]
where $\Gamma(a,b)$ denotes the Gamma distribution 
with mean $a/b$ and $\textrm{Exp}(b)$ denotes the Exponential distribution 
with mean $1/b$). For simplicity, we fixed the initial latent states at their true values. We 
performed a pilot run of the 
PMMH scheme with 50 particles to give an approximate covariance matrix $\widehat{\textrm{Var}}(c)$ 
and approximate posterior mean $\hat{c}$. By performing further pilot runs we found that using 
250 particles gave the variance in the noise in the estimated log-posterior as 1.54. We 
therefore took $N=250$ particles for the main monitoring runs, which typically consisted of $2\times 10^5$ iterations 
of each scheme, with the $\log(c_{i})$ updated in a single block using a Gaussian random walk 
proposal kernel. We used an innovation variance matrix given 
by $\lambda\frac{2.38^{2}}{3}\widehat{\textrm{Var}}(c)$. For PMMH, further pilot runs were performed 
to determine an appropriate scaling $\lambda$. We used $\lambda=0.6$ (which gave an acceptance 
rate of around $8\%$) for the main run. The cost of computing an approximation to the marginal likelihood (under the LNA) 
versus computing an estimate of marginal likelihood under the MJP scales roughly as $1:780$ for LNA : MJP and we might therefore 
expect that a larger value of $\lambda$ will be optimal for daPMMH-LNA. In order to investigate effect of $\lambda$ on the daPMMH-LNA scheme, 
we report results for $\lambda=0.6,1,2,3,4$.

\begin{figure*}
\centering
\psfrag{log(gr)}[][][1.2][0]{$\log(\gamma_{R})$}
\psfrag{log(gp)}[][][1.2][0]{$\log(\gamma_{P})$}
\psfrag{log(kp)}[][][1.2][0]{$\log(\kappa_{P})$}
\psfrag{log(b0)}[][][1.2][0]{$\log(b_0)$}
\psfrag{log(b1)}[][][1.2][0]{$\log(b_1)$}
\psfrag{log(b2)}[][][1.2][0]{$\log(b_2)$}
\psfrag{log(b3)}[][][1.2][0]{$\log(b_3)$}
\psfrag{log(sig)}[][][1.2][0]{$\log(\sigma)$}
\includegraphics[angle=270,width=\textwidth]{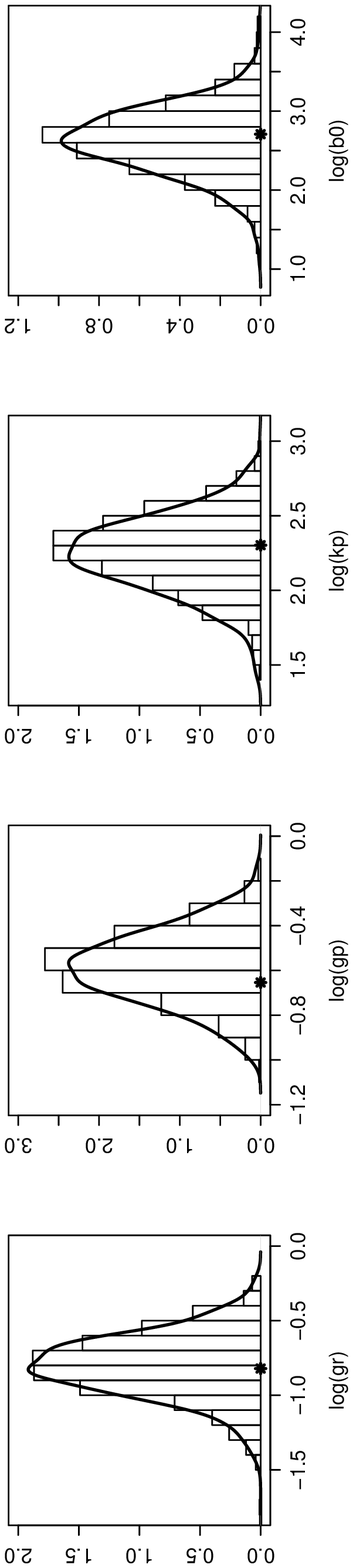}
\includegraphics[angle=270,width=\textwidth]{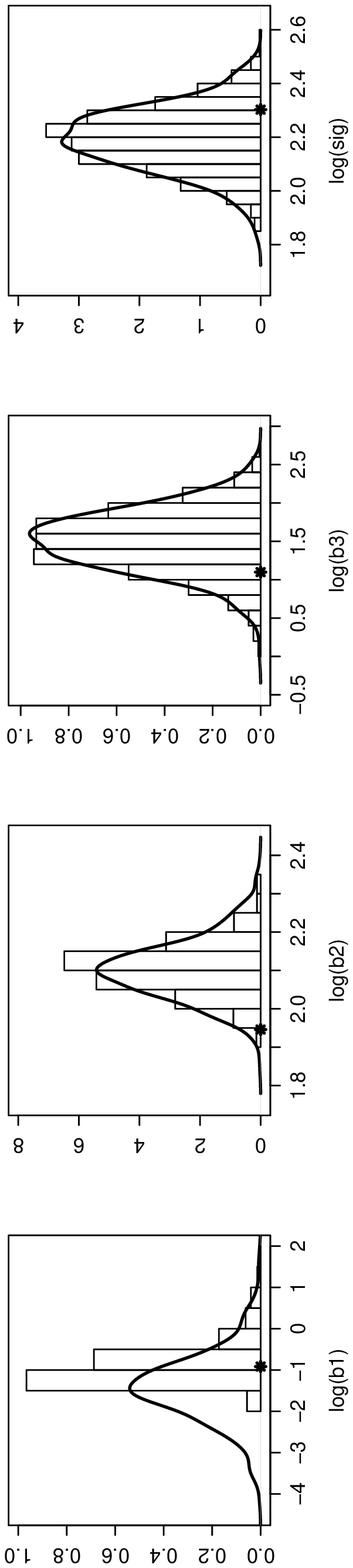}
\caption{Gene expression model. Marginal posterior distributions under the MJP (histograms) and 
LNA (solid line). True values of each $\log(c_i)$ are indicated ($\ast$).}\label{fig:fig0}
\end{figure*}

\begin{figure*}
\centering
\psfrag{Logpa(D1|c)}[][][1.0][0]{$\log(p_{a}(\mathbf{y}|c))$}
\psfrag{Logp(D1|c)}[][][1.0][0]{$\log(\widehat{p}(\mathbf{y}|c))$}
\includegraphics[width=7.5cm,height=7.5cm,angle=270]{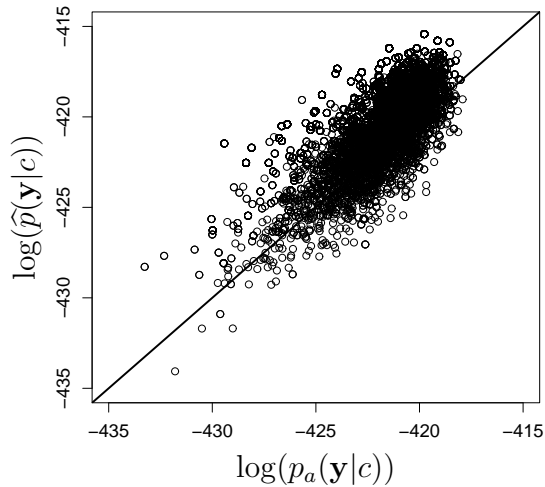}
\caption{Log-marginal likelihood estimates $\log(\widehat{p}(\mathbf{y}|c))$ under the MJP against 
the corresponding log-marginal likelihood estimate under the LNA, using $10,000$ values of $c$ 
sampled from the posterior $p(c|\mathbf{y})$ for the gene expression model.}\label{fig:fig3}
\end{figure*}

\begin{table*}[t]
	\begin{center}
	\begin{tabular}{|l|ccccc|}
	\hline	
Algorithm & $\alpha_{1}$ & $\alpha_{2|1}$ & CPU time (s) & ESS$_{min}$ &  Rel. ESS$_{min}$/s \\
\hline
PMMH                       &0.077 &1.000 &350657 &524 &1.00    \\
daPMMH-LNA ($\lambda=0.6$) &0.218 &0.198 &77704 &316 & 2.72    \\
daPMMH-LNA ($\lambda=1$)   &0.137 &0.178 &50840 &394 & 5.19    \\
daPMMH-LNA ($\lambda=2$)   &0.051 &0.163 &20155 &246 &8.18   \\
daPMMH-LNA ($\lambda=3$)   &0.029 &0.149 &11667 &153 &8.76    \\
daPMMH-LNA ($\lambda=4$)   &0.023 &0.182 &9518  &120 & 8.44    \\
	\hline
	\end{tabular}
	\caption{Gene expression model. Stage 1 acceptance rate $\alpha_{1}$, Stage 2 acceptance rate $\alpha_{2|1}$, 
CPU time (to the nearest second), minimum effective sample size (ESS$_{min}$, to the nearest whole number) and 
minimum effective sample size per second, relative to the corresponding value obtained from the vanilla PMMH scheme. All values are 
based on $2\times 10^5$ iterations.}\label{tab:tab3}
	\end{center}
\end{table*}

Figure~\ref{fig:fig0} summarises the output of the PMMH scheme which we find to be consistent 
with the output of the daPMMH-LNA scheme (not reported). We also give kernel density estimates 
of the marginal parameter posteriors under the LNA. The posterior samples appear to be consistent with 
the true values that produced the data although we see some discrepancy between the 
LNA and MJP posteriors. Table~\ref{tab:tab3} shows Stage 1 acceptance rate $\alpha_{1}$, 
Stage 2 acceptance rate $\alpha_{2|1}$, 
CPU time, minimum (over the parameters) effective sample size (ESS$_{min}$) and minimum effective sample size per second, 
relative to the corresponding value obtained from the PMMH scheme. The effect of increasing the scaling 
parameter $\lambda$ (which in turn increases the innovation variance for the Gaussian random walk update) 
can clearly be seen. When $\lambda=3$ we see 
an 8 fold improvement in overall efficiency (as measured by relative ESS$_{min}$ per second). The result 
is relatively robust to the choice of $\lambda$, with a relative ESS$_{min}$ per second of 2.72 when using the 
same scaling as PMMH ($\lambda=0.6$). 

The accuracy of the LNA can be assessed through inspection of Figures~\ref{fig:fig0} and \ref{fig:fig3}. There is a 
noticeable discrepancy in the marginal posteriors for $\log(b_{1})$ and $\log(b_{2})$. Despite this, Figure~\ref{fig:fig3} 
suggests that the LNA provides a reasonable 
approximation to the MJP in regions of high posterior density, and we recorded an empirical Stage 2 acceptance 
probability of around 0.18.

\subsection{Epidemic model}\label{epi}
Finally, we consider a Susceptible--Infected--Removed (SIR) epidemic model involving two species 
(susceptibles $\mathcal{X}_{1}$ and infectives $\mathcal{X}_{2}$) and two reaction channels (infection of 
a susceptible and removal of an infective):
\begin{align*}
\mathcal{R}_1:\quad \mathcal{X}_{1}+\mathcal{X}_{2} &\xrightarrow{\phantom{a}\beta\phantom{a}} 2\mathcal{X}_{2}\\
\mathcal{R}_2:\quad \mathcal{X}_{2} &\xrightarrow{\phantom{a}\gamma\phantom{a}} \emptyset.
\end{align*}
The system can be seen as a special case of the Lotka-Volterra system with $c_{1}=0$. We let $c=(\beta,\gamma)'$ 
denote the unknown parameter vector. The stoichiometry matrix is given by
\[
S = \left(\begin{array}{rr} 
-1 & 0\\
 1 & -1
\end{array}\right)
\]
and the associated hazard function is 
\[
h(X,c) = (\beta X_{1}X_{2}, \gamma X_{2})'
\]
where $X=(X_{1},X_{2})'$ denotes the state of the system 
at time $t$. For the linear noise approximation, the Jacobian 
matrix $F_t$ is given by
\[
F_{t} = \left(\begin{array}{cc} 
-\beta z_{2} & -\beta z_{1} \\
\beta z_{2} & \beta z_{1}-\gamma 
\end{array}\right)
\]  
where $z=(z_1,z_2)'$ is the state at time $t$ of the deterministic 
process satisfying (\ref{LNA1}).

We consider the Abakaliki small pox dataset given in \cite{bailey1975} and studied by 
\cite{oneill1999}, \cite{fearnhead2004} and \cite{boys2007} among others. Page 125 of 
\cite{bailey1975} provides a complete set of 29 inter-removal times, measured in days, 
from a smallpox outbreak in a community of 120 individuals in Nigeria. We report the data here 
as the days on which the removal of individuals actually took place, with the first day set to 
be time 0 (Table~\ref{tab:tab4}). We assume an SIR model for the data with observations being 
equivalent to daily measurements of $X_{1}+X_{2}$ (as there is a fixed population size). In addition, 
and for simplicity, we assume that a single individual remained infective just after the first 
removal occurred. We analyse the data under the assumption of no measurement error. This assumption 
can be incorporated into the PMMH algorithm by calculating the un-normalised weight in step 2(b) 
of the SMC scheme as
\[
w_{t+1}^{*i}=\left\{\begin{array}{cc}
1,\quad & x_{t+1}^{i}=y_{t+1}\\
0,\quad & \textrm{otherwise}
\end{array}\right.
\] 
The marginal likelihood under the LNA can be computed using the algorithm described in 
\ref{A2} with $G'=(1,1)$ and $\Sigma=0$. Note that for this example, the cost of computing 
the LNA marginal likelihood versus an estimate of marginal likelihood under the MJP 
scales roughly as $1:34$ for LNA : MJP.

\begin{table*}[t]
	\begin{center}
	\begin{tabular}{|l|llllllllllll|}
	\hline	
Day & 0 & 13 & 20 & 22 & 25 & 26 & 30 &35 &38 & 40 & 42 &47\\
No. of removals & 1 & 1 & 1 & 1 & 3  & 1 & 1 & 1 & 1 & 2 & 2 &1\\ 
	\hline
Day & 50 & 51 & 55 & 56 & 57 & 58 & 60 &61 &66 & 71 & 76 & \\
No. of removals & 1 & 1 & 2 & 1 & 1  & 1 & 2 & 1 & 2 & 1 & 1 & \\ 
\hline
	\end{tabular}
	\caption{Abakaliki smallpox data.}\label{tab:tab4}
	\end{center}
\end{table*}
 
We followed \cite{fearnhead2004} by taking $\beta \sim \Gamma(10,10^4)$ and 
$\gamma\sim \Gamma (10,10^2)$ \emph{a priori}. A pilot run of the 
PMMH scheme with 500 particles was used to give an approximate covariance matrix $\widehat{\textrm{Var}}(c)$ 
and approximate posterior mean $\hat{c}$. By performing further pilot runs we found that using 
2000 particles gave the variance in the noise in the estimated log-posterior as 1.25. We 
therefore took $N=2000$ particles for the main monitoring runs, which typically consisted of $10^5$ iterations 
of each scheme, with the $\log(c_{i})$ updated in a single block using a Gaussian random walk 
proposal kernel with innovation variance $\lambda\frac{2.38^{2}}{3}\widehat{\textrm{Var}}(c)$. For PMMH, 
a number of short pilot runs suggested that $\lambda=1.1$ (which gave an acceptance rate of 0.23) was 
close to optimal. 

\begin{figure*}
\centering
\psfrag{log(be)}[][][1.2][0]{$\log(\beta)$}
\psfrag{log(ga)}[][][1.2][0]{$\log(\gamma)$}
\psfrag{MJP}[][][1.2][0]{MJP}
\psfrag{LNA}[][][1.2][0]{LNA}
\includegraphics[angle=270,width=\textwidth]{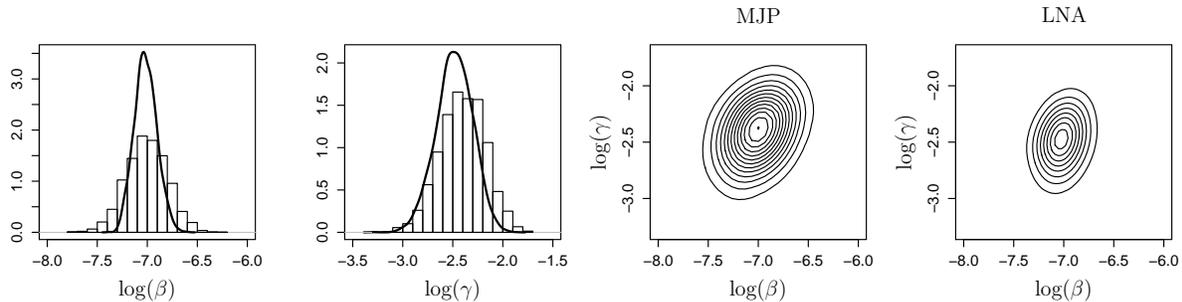}
\caption{Epidemic model. Marginal posterior distributions under the MJP (histograms) and 
LNA (solid line), and contour plots of the joint posterior under the MJP (left) and LNA (right).}\label{fig:figE}
\end{figure*}

\begin{table*}[t]
	\begin{center}
	\begin{tabular}{|l|ccccc|}
	\hline	
Algorithm & $\alpha_{1}$ & $\alpha_{2|1}$ & CPU time (s) & ESS$_{min}$ &  Rel. ESS$_{min}$/s \\
\hline
PMMH                       &0.226 &1.000 &4981 &7469 &1.00    \\
daPMMH-LNA ($\tau=1$, $\lambda=1.1$) &0.252 &0.402 &1208 &1478 &0.82    \\
daPMMH-LNA ($\tau=4$, $\lambda=1.1$)   &0.345 &0.509 &1634 &3444 &1.41    \\
daPMMH-LNA ($\tau=5$, $\lambda=1.1$)   &0.358 &0.499 &1698 &4374 &1.72   \\
daPMMH-LNA ($\tau=6$, $\lambda=1.1$)   &0.372 &0.488 &1763 &2831 &1.07    \\
daPMMH-LNA ($\tau=5$, $\lambda=3$)     &0.180 &0.476 &890  &2920 &2.19    \\
daPMMH-LNA ($\tau=5$, $\lambda=4$)     &0.144 &0.471 &762  &2471 &2.16    \\
daPMMH-LNA ($\tau=5$, $\lambda=5$)     &0.120 &0.468 &649  &2008 &2.06    \\
	\hline
	\end{tabular}
	\caption{Epidemic model. Stage 1 acceptance rate $\alpha_{1}$, Stage 2 acceptance rate $\alpha_{2|1}$, 
CPU time (to the nearest second), minimum effective sample size (ESS$_{min}$, to the nearest whole number) and 
minimum effective sample size per second, relative to the corresponding value obtained from the vanilla PMMH scheme. All values are 
based on $10^5$ iterations.}\label{tab:tabE1}
	\end{center}
\end{table*}

Figure~\ref{fig:figE} gives marginal posterior densities under the MJP (using the output of the PMMH scheme) 
and the LNA (using the output of the daPMMH-LNA scheme without Stage 2). We see that the LNA substantially 
underestimates the uncertainty in $\beta$. Use of the LNA as a surrogate model in this case will likely lead 
to rejected parameter draws at Stage 1 that would otherwise be accepted at Stage 2. We alleviate this problem 
by scaling the log marginal likelihood under the LNA by an amount $1/\tau$, where $\tau$ is chosen to 
maximise the efficiency of the delayed acceptance scheme. Specifically, we replace 
$p_{a}(\mathbf{y}|c)$ in Algorithm~1 with $p_{a}(\mathbf{y}|c)^{\frac{1}{\tau}}$. To determine an appropriate value 
for $\tau$, we fixed the scaling $\lambda$ at 1.1 and ran the daPMMH-LNA scheme for $\tau$ in the range $[1,10]$. 
Table~\ref{tab:tabE1} reports results for $\tau\in\{1,4,5,6\}$. We see that as $\tau$ increases so does the 
Stage 1 acceptance rate, resulting in an increase in CPU time (as the expensive MJP simulator is run more often). 
However, the Stage 2 acceptance rate also increases, suggesting an optimal value of $\tau$. We found that $\tau=5$ 
is optimal for the range considered. We therefore fixed $\tau=5$ and varied the scaling $\lambda$. For $\lambda\in
\{3,4,5\}$ it is possible to achieve a 2-fold increase in efficiency over PMMH.

\section{Discussion}\label{conc}
We have proposed two delayed acceptance Particle Mar-ginal Metropolis-Hastings algorithms, analogues of the 
delayed acceptance Metropolis-Hastings scheme of \cite{ChristenF05}. We have shown that both lead to 
a chain with the desired stationary distribution and applied them to the problem of parameter estimation in 
Markov jump processes with state-dependent rate parameters. In both analogues the true posterior that is used 
in \cite{ChristenF05} is replaced with an unbiased approximation obtained through a particle filter. In 
the second analogue the fast deterministic approximation is replaced with a relatively fast stochastic 
approximation that is also obtained via a particle filter. The need for such an approach is motivated 
by the potentially huge computational cost of performing particle MCMC for the MJP directly, 
where each iteration requires implementation of a particle filter with $N$ particles, and a 
complete run of the stochastic simulation algorithm is required for each particle. 

The delayed acceptance PMMH scheme aims to avoid calculating an estimate of marginal likelihood 
(and therefore running the particle filter) under the MJP for proposals that are likely to 
be rejected, by implementing a preliminary screening step that uses a cheap approximation of 
the marginal likelihood. We explored two approximations, the chemical Langevin equation (CLE) and 
the linear noise approximation (LNA). The LNA can be viewed as an approximation to the CLE. 
Thus, providing the Euler time-step is not too large the CLE leads to a greater effective 
sample size over a fixed number of iterations. However under Gaussian observation regimes 
the marginal likelihood under the LNA is tractable, whereas the marginal likelihood under 
the CLE is generally intractable whatever the observation regime. We therefore replaced the 
true posterior under the CLE approximation with a stochastic approximation to this, also 
obtained via a particle filter.  We tested both schemes on a Lotka-Volterra system where the 
observed counts follow a Poisson distribution with expectation equal to the true count. We showed how the 
LNA can be used to obtain a reasonable deterministic approximation to the marginal likelihood 
even though the observations are not Gaussian and created a scheme which is approximately an 
order of magnitude more efficient than the standard PMMH scheme. Even though the particle 
filter is computationally much more costly than simply integrating the LNA, using the CLE 
we are still able to double the efficiency compared with the standard PMMH scheme. In a 
further application of the LNA scheme to a more complex MJP, with a larger number of 
unknown parameters we again obtained a speed up of approximately an order of magnitude.

The proposed methodology can in principle be applied to any stochastic kinetic model and in Section~\ref{epi} we 
applied the delayed acceptance scheme (using the LNA) to a simple epidemic model. For this example, we found that 
an estimate of marginal likelihood under the MJP could be computed relatively cheaply. In spite of this, running the 
delayed acceptance scheme is still worthwhile, and we observed an overall increase in efficiency of at least a factor 
of two.   

The efficiency of both proposed delayed acceptance PMMH schemes can be improved in 
a number of ways. Both schemes can be parallelised and will 
benefit from recent work on the use of graphics cards for Monte Carlo methods 
\citep{Lee10}. In addition, in high 
signal-to-noise scenarios, the variance of the marginal likelihood estimator under both the CLE 
and MJP could be reduced through implementation of an auxiliary particle filter such as that considered 
by \cite{pitt12}. The interplay between the number of particles, and choice of scaling for the 
RWM proposal, and the efficiency of the scheme is non-trivial. For example, increasing the 
number of particles increases the CPU time per iteration but (e.g. \cite{andrieu09b}) 
should lead to a more efficient PMMH algorithm in terms of ESS for a fixed number of iterations. 
However with a delayed acceptance algorithm we might expect less of an increase in ESS once the 
accuracy of the stochastic approximation exceeds that of the deterministic approximation since the 
Stage 2 acceptance rate depends on the ratio of these. Our tuning of the algorithms was 
relatively ad hoc; with sound tuning advice driven by theory it is possible that further 
efficiency gains might be obtained.

Our demonstration of detailed balance showed that when a stochastic
estimate of the marginal likelihood is used at Stage 1 as well as
Stage 2, the independence of
the estimators is unnecessary. This suggests that a positive correlation
between the two might increase the Stage 2 acceptance rate;
unfortunately it was not obvious how to achieve this for our
particular estimators. It is also straightforward to extend our
derivation to apply to a $k$-Stage delayed acceptance algorithm, using
 a sequence of $k-1$ approximations. Such a sequence would need
 a careful design as the increase in accuracy at each
 stage would need to outweigh the increase in computational cost, and
  we do not pursue this here.   

\appendix

\section{Appendix}
Recall that $\mathbf{x}=\{x_{t}\,|\, 1\leq t \leq T\}$ denotes values of the latent MJP 
and $\mathbf{y}=\{y_{t}\,|\, t=1,2,\ldots ,T\}$ denotes the collection of (noisy) observations 
on the MJP at discrete times. In addition, we define $\mathbf{x}_{t}=\{x_{s}\,|\, t-1<s\leq t\}$ 
and $\mathbf{y}_{t}=\{y_{s}\,|\, s=1,2,\ldots, t\}$.
\subsection{PMMH scheme}\label{A}

The PMMH scheme has the following algorithmic form.
\begin{enumerate}
\item Initialisation, $i=0$,
\begin{itemize}
\item[(a)] set $c^{(0)}$ arbitrarily and
\item[(b)] run an SMC scheme targeting $p(\mathbf{x}|\mathbf{y},c^{(0)})$, 
and let $\widehat{p}(\mathbf{y}|c^{(0)})$ denote the marginal likelihood estimate
\end{itemize}
\item For iteration $i\geq 1$,
\begin{itemize}
\item[(a)] sample $c^{*}\sim q(\cdot | c^{(i-1)})$,
\item[(b)] run an SMC scheme targeting $p(\mathbf{x}|\mathbf{y},c^{*})$, and let $\widehat{p}(\mathbf{y}|c^{*})$ 
denote the marginal likelihood estimate, 
\item[(c)] with probability min$\{1,A\}$ where
\[
A=\frac{\widehat{p}(\mathbf{y}|c^{*}) p(c^{*})}{\widehat{p}(\mathbf{y}|c^{(i-1)}) p(c^{(i-1)})}
\times \frac{q(c^{(i-1)} | c^{*})}{q(c^{*} | c^{(i-1)})}
\]
accept a move to $c^{*}$ otherwise store the current values
\end{itemize}
\end{enumerate}
Note that the PMMH scheme can be used 
to sample the joint posterior 
$p(c,\mathbf{x}|\mathbf{y})$. Essentially, a proposal mechanism of the form 
$q(c^{*}|c)\widehat{p}(\mathbf{x}^{*}|\mathbf{y},c^{*})$, 
where $\widehat{p}(\mathbf{x}^{*}|\mathbf{y},c^{*})$ is an SMC 
approximation of $p(\mathbf{x}^{*}|\mathbf{y},c^{*})$, is used. 
The resulting MH acceptance ratio is as above. Full details of the PMMH scheme 
including a proof establishing that the method leaves the target 
$p(c,\mathbf{x}|\mathbf{y})$ invariant can be found in \citet{andrieu10}.

\subsection{SMC scheme}\label{A1}

A sequential Monte Carlo estimate of the marginal likelihood $p(\mathbf{y}|c)$ under the MJP 
can be constructed using (for example) the bootstrap filter of \cite{gordon93}. Algorithmically, 
we perform the following sequence of steps.
\begin{enumerate}
\item Initialisation.
\begin{itemize}
\item[(a)] Generate a sample of size $N$, $\{x_{1}^{1},\ldots ,x_{1}^{N}\}$ from the initial density $p(x_{1})$.
\item[(b)] Assign each $x_{1}^{i}$ a (normalised) weight given by
\[
w_{1}^{i}=\frac{w_{1}^{*i}}{\sum_{i=1}^{N}w_{1}^{*i}}, \quad\textrm{where}\quad w_{1}^{*i}=p(y_{1}|x_{1}^{i},c)\,.
\]
\item[(c)] Construct and store the currently available estimate of marginal likelihood,
\[
\widehat{p}({y}_{1}|c) = \frac{1}{N}\sum_{i=1}^{N} w_{1}^{*i}\,.
\]
\item[(d)] Resample $N$ times with replacement from $\{x_{1}^{1},\ldots ,x_{1}^{N}\}$ 
with probabilities given by $\{w_{1}^{1},\ldots ,w_{1}^{N}\}$.
\end{itemize}
\item For times $t=1,2,\ldots ,T-1$,
\begin{itemize}
\item[(a)] For $i=1,\ldots ,N$: draw $\mathbf{X}_{t+1}^{i}\sim p\big(\mathbf{x}_{t+1}|{x}_{t}^{i},c\big)$ using the Gillespie algorithm. 
\item[(b)] Assign each $\mathbf{x}_{t+1}^{i}$ a (normalised) weight given by
\[
w_{t+1}^{i}=\frac{w_{t+1}^{*i}}{\sum_{i=1}^{N}w_{t+1}^{*i}}, \quad\textrm{where}\quad w_{t+1}^{*i}=p(y_{t+1}|x_{t+1}^{i},c)\, .
\]
\item[(c)] Construct and store the currently available estimate of marginal likelihood,
\begin{align*}
\widehat{p}(\mathbf{y}_{t+1}|c) &= \widehat{p}(\mathbf{y}_{t}|c)\widehat{p}(y_{t+1}|\mathbf{y}_{t},c)\\
&=\widehat{p}(\mathbf{y}_{t}|c)\frac{1}{N}\sum_{i=1}^{N} w_{t+1}^{*i}\,.
\end{align*}
\item[(d)] Resample $N$ times with replacement from $\{\mathbf{x}_{t+1}^{1},\ldots ,\linebreak[1]\mathbf{x}_{t+1}^{N}\}$ 
with probabilities given by $\{w_{t+1}^{1},\ldots ,w_{t+1}^{N}\}$.
\end{itemize}
\end{enumerate}

\subsection{Marginal likelihood under the linear noise approximation}\label{A2}
Assume an observation regime of the form
\[
Y_{t}=G'X_{t}+\varepsilon_{t}\,,\qquad \varepsilon_{t}\sim \textrm{N}\left(0,\Sigma\right)
\] 
where $G$ is a constant matrix of dimension $u\times p$ and $\varepsilon_{t}$ is a length-$p$ 
Gaussian random vector.

Now suppose that 
$X_{1}\sim N(a,C)$ \emph{a priori}. The marginal likelihood 
under the LNA, $p_{a}(\mathbf{y}|c)$ can be obtained as follows.
\begin{enumerate}
\item Initialisation. Compute 
\[
p_{a}(y_{1}|c)=\phi\left(y_{1}\,;\, G'a\,,\,G'CG+\Sigma\right)
\]
where $\phi(\cdot\,;\,a\,,\,C)$ denotes the Gaussian density with 
mean vector $a$ and variance matrix $C$. The posterior at time $t=1$ is therefore 
$X_{1}|y_{1}\sim N(a_{1},C_{1})$ where
\begin{align*}
a_{1} &= a+CG\left(G'CG+\Sigma\right)^{-1}\left(y_{1}-G'a\right) \\
C_{1} &= C-CG\left(G'CG+\Sigma\right)^{-1}G'C\,.
\end{align*}
 
\item For times $t=1,2,\ldots ,T-1$,
\begin{itemize}
\item[(a)] Prior at $t+1$. Initialise the LNA with $z_{t}=a_{t}$, $m_{t}=0$ and $V_{t}=C_{t}$. 
Note that this implies $m_{s}=0$ for all $s>t$. Therefore, integrate the ODEs (\ref{LNA1}) and (\ref{odeV}) 
forward to $t+1$ to obtain $z_{t+1}$ and $V_{t+1}$. Hence
\[
X_{t+1}|\mathbf{y}_{t}\sim N(z_{t+1},V_{t+1})\,.
\]
\item[(b)] One step forecast. Using the observation equation, we have that 
\[
Y_{t+1}|\mathbf{y}_{t}\sim N\left(G'z_{t+1},G'V_{t+1}G+\Sigma\right)\,.
\]
Compute
\begin{align*}
p_{a}(\mathbf{y}_{t+1}|c)&=p_{a}(\mathbf{y}_{t}|c)p_{a}(y_{t+1}|\mathbf{y}_{t},c)\\
&=p_{a}(\mathbf{y}_{t}|c)\,\phi\left(y_{t+1}\,;\, G'z_{t+1}\,,\,G'V_{t+1}G+\Sigma\right)\,.
\end{align*}
\item[(c)] Posterior at $t+1$. Combining the distributions in (a) and (b) gives the joint 
distribution of $X_{t+1}$ and $Y_{t+1}$ (conditional on $\mathbf{y}_{t}$ and $c$) as
\[
\left(\begin{array}{c}
	X_{t+1} \\	
	Y_{t+1}
	\end{array}\right)\sim N\left\{\left(\begin{array}{c}
	z_{t+1} \\
	G'z_{t+1} 	
	\end{array} \right )\,,\, \left(\begin{array}{cc}
	V_{t+1} & V_{t+1}G  \\
	G'V_{t+1} & G'V_{t+1}G+\Sigma  	 
	\end{array}   \right ) \right \} 
\]
and therefore $X_{t+1}|\mathbf{y}_{t+1}\sim N(a_{t+1},C_{t+1})$ where
\begin{align*}
a_{t+1} &= z_{t+1}+V_{t+1}G\left(G'V_{t+1}G+\Sigma\right)^{-1}\left(y_{t+1}-G'z_{t+1}\right) \\
C_{t+1} &= V_{t+1}-V_{t+1}G\left(G'V_{t+1}G+\Sigma\right)^{-1}G'V_{t+1}\,.
\end{align*}

\end{itemize}
\end{enumerate}


\bibliographystyle{apalike}
\bibliography{refs}   

\end{document}